\documentclass[final,3p,times,onecolumn,authoryear]{elsarticle}

\usepackage{amssymb}
\usepackage{amsmath}
\usepackage{array}
\usepackage{booktabs}
\usepackage{caption}
\usepackage{makecell}
\usepackage{hyperref}
\usepackage{placeins}
\usepackage{multirow}
\usepackage{comment}
\usepackage{graphicx}
\usepackage{subcaption}
\usepackage{xurl}

\begin{document}

\begin{frontmatter}

\title{Demystifying the trend of the healthcare index: Is historical price a key driver?}

\author[aff1]{Payel Sadhukhan}
\author[aff2]{Samrat Gupta\corref{cor1}}
\author[aff3]{Subhasis Ghosh}
\author[aff4,aff5]{Tanujit Chakraborty}

\address[aff1]{Army Institute of Management, Kolkata, India}
\address[aff2]{Indian Institute of Management, Ahmedabad, India}
\address[aff3]{ICFAI University, Tripura, India}
\address[aff4]{SAFIR, Sorbonne University, Abu Dhabi, UAE}
\address[aff5]{Sorbonne Center for AI, Paris, France}

\cortext[cor1]{samratg@iima.ac.in}

\begin{abstract}
Healthcare sector indices consolidate the economic health of pharmaceutical, biotechnology, and healthcare service firms. The short-term movements in these indices are closely intertwined with capital allocation decisions affecting research and development investment, drug availability, and long-term health outcomes. This research investigates whether historical open–high–low–close (OHLC) index data contain sufficient information for predicting the directional movement of the opening index on the subsequent trading day. The problem is formulated as a supervised classification task involving a one-step-ahead rolling window. A diverse feature set is constructed, comprising original prices, volatility-based technical indicators, and a novel class of nowcasting features derived from mutual OHLC ratios. The framework is evaluated on data from healthcare indices in the U.S. and Indian markets over a five-year period spanning multiple economic phases, including the COVID-19 pandemic. The results demonstrate robust predictive performance, with accuracy exceeding 0.8 and Matthews correlation coefficients above 0.6. Notably, the proposed nowcasting features have emerged as a key determinant of the market movement. We have employed the Shapley-based explainability paradigm to further elucidate the contribution of the features: outcomes reveal the dominant role of the nowcasting features, followed by a more moderate contribution of original prices. This research offers a societal utility: the proposed features and model for short-term forecasting of healthcare indices can reduce information asymmetry and support a more stable and equitable health economy.
\end{abstract}


\end{frontmatter}



\section{Introduction}
\label{intro}
A healthcare index is an aggregate of stocks from leading drug manufacturers, healthcare companies, and biotechnology firms. While its fluctuations primarily interest traders and investors tracking financial markets, the index holds broader relevance for societal well-being \citep{pharma-growth1}. The rise or fall of this class of indices serves as a key indicator for real-time understanding of society's well-being at large, reflecting and influencing a range of diverse aspects, including public health outcomes, economic consequences, and national security and ethics \citep{pharma-society}. 

It is a powerful indicator and influencer of the state's and the public's interest in medical innovation, competency, and societal health \citep{pharma-society2,pharma-society3}. A sustained upward trend is often rooted in successful clinical trial results \footnote{\url{https://www.ainvest.com/news/eli-lilly-stock-soars-10-5-positive-diabetes-drug-trial-market-uncertainty-2508/}}, regulatory approvals for new drugs \footnote{\url{https://brandfinance.com/press-releases/covid-19-vaccine-race-boosts-brand-values-across-pharma-giants}} \citep{new-drug}, or breakthroughs in treating widespread or life-threatening diseases \footnote{\url{https://www.investopedia.com/s-and-p-500-gains-and-losses-today-apple-stock-climbs-eli-lilly-slumps-after-disappointing-trial-results-11787193}} \citep{pharma-rise}. The persistence of growth can also be attributed to the stability of the economic policies underpinning the domain \citep{pharma-economic1,pharma-economic2}.

The understanding of the rise and fall of healthcare index, whether national or global, requires deciphering of two distinct facets: i] long-term events and ii] short-term fluctuations. Long-term events typically stem from foundational changes in the sector's landscape, including events such as drug approvals, landmark economic policy shifts, new disease prevalence, or major patent expirations. The onset of this class of events and their impact on the index can be anticipated through a qualitative understanding of the demographic and sectoral aspects \citep{pharma-economic1}, providing stakeholders with an extended timeframe for strategic response. Conversely, the short-term facets are predominantly driven by day-to-day index movements and fluctuations over a time period. Having a semantical interpretation of these movements and their causation is notably more complex, as they are influenced by a confluence of multidimensional aspects, such as investor sentiment, market reactions, and liquidity conditions \citep{short-term1, short-term2}. 

Given this critical yet challenging nature of short-term fluctuations, we focus on deciphering the randomness in the above-stated \emph{noise element}. Table~\ref{tab:vol} shows the average diurnal volatilities of different sectoral indices in the USA and the Indian market. The technical formulations for calculating volatility can be found in ~\ref{sec:vol}. Cyclical sectors, such as banks and the automobile industry, exhibit volatile swings with economic booms and busts \citep{cyclical-stock1,cyclical-stock2}. In contrast, healthcare indices exhibit lower volatility due to the ever-lasting requirement for medicines and associated supplies \citep{pharma-predictibility}. This inherent stability suggests that the movement of healthcare indices may contain a predictable pattern and is camouflaged by the apparent chaos of the daily movement. Therefore, instead of seeking a semantic interpretation of fluctuations in the long-term movements, we aim to model the movement of the healthcare index by estimating the information contained in the short-term movement of the raw price indices. 

\begin{table*}[htbp]
  \centering
  \caption{Sectoral volatilities in USA and Indian markets}
  \label{tab:vol}
  \begin{tabular}{|c|c|c|}
    \toprule
    \textbf{Sectors} & {U.S.A.}& {India}\\
    & \textbf{Close index} & \textbf{Close index} \\
    \midrule
    Finance & 0.60 & 0.56 \\
    IT  & 22.06 & 0.52 \\
    Oil and Gas  & 0.53  & 0.53 \\
    Healthcare  & 0.42 & 0.39 \\
    \bottomrule
  \end{tabular}
  
  \smallskip
\end{table*}

Machine Learning (ML) possesses the capability to understand deep dependencies and patterns among data aspects and is often used in forecasting \citep{economic-forecasting}. ML helps in discovering latent dependencies that may otherwise go unnoticed in human perception \citep{machine-learning}. The low volatility and non-cyclicality of the healthcare sector lead to reliable and less random patterns \citep{pharma-stationarity}. These patterns can be efficiently recognized by ML methods and subsequently utilized to make efficient predictions of upcoming trends \citep{pharma-trend1}. The veracity of predictions can aid in investment-related decision-making, guiding investors on when to invest in stocks from the healthcare domain \citep{ml-decision}. The development of an ML model that can provide accurate insight into the rise or fall of the next day's open index for the healthcare sector will have high utility for traders investing in this domain.

The primary research question addressed in this work is --- \emph{Can we find out the key drivers of healthcare indices and build interpretable ML models to predict the rise or fall in healthcare indices?} 
\\ The close intertwining between this sector's index movement and societal developments, along with the diminished cyclicality of the sector, motivates us to develop an ML model for capturing the association between the historical indices and the upcoming rise or fall of the trend in this sector \citep{pharma-defensive}. 
Our methodology begins by deriving the target class, \emph{rise} or \emph{fall}. This is obtained by computing the difference between the open index of the upcoming trading day and the indices of the current trading day. Although the open index for the upcoming trading day is unknown in practice, we construct a labeled dataset from historical records, thereby framing a supervised classification problem. 
A model's performance is fundamentally rooted in its data. For a model to perform well, careful feature engineering is required; features must associate strongly with the target class -- in this case, the rise/fall of the indices. To this end, we explore the role of historical momentum as well as current day market context.
We begin by including the raw OHLC prices of the current trading day in the feature set. In addition, we engineer two complementary classes of features from the raw prices as well. One class of these features focuses on raw prices of multiple previous days and includes volatility-focused parameters trusted by traders, namely Bollinger Bands, Keltner Channels, and Donchian Channels. These indicators are built on the statistical summarization of historical data, inherently introducing a lag \citep{volatility-lag}. They provide glimpses of past market behavior, but there is no near-real-time snapshot of current market conditions. To address this limitation and provide a near-real-time snapshot, we design a new class of features. It comprises the mutual ratios of the current day's index set and serves as a nowcasting tool in the working features set. The ratios provide a detailed picture of the current day's market movement and serve as a vital clue for predicting the next day's open index. In essence, the feature set consists of complementary market signals at different temporal scales.

Data source is a foundational component of any meaningful analysis \citep{data-quality}. In this research, we have utilized sectoral data from the healthcare domains of the US and Indian Markets. The daily indices (open, close, high, and low) from April 1, 2019, to March 31, 2024, are considered for both countries. This five-year period spans diverse market phases, including the COVID-19 pandemic, thereby providing a comprehensive test of our models. The choice of these two markets is deliberate and is included to enhance the generalizability of our findings. The U.S. healthcare market, valued at approximately USD 640 billion in 2024, is largely driven by breakthrough R$\&$D innovations, company mergers, and patent-protected drugs. In contrast, the Indian market, valued at approximately 65 billion the same year, is dominated by generic drug production, rising healthcare needs, lifestyle diseases, and volume-driven sales. These structurally divergent aspects allow us to identify cross-market trends, volatility patterns, and macroeconomic linkages that are crucial for validating the ML models trained on them. An array of diversified classifiers is employed in this study to decipher the relationship between the engineered features and the target class. We have evaluated the correctness of predictions using \emph{accuracy} and \emph{Mathews Correlation Coefficient (MCC) scores}. The latter, in particular, provides a balanced measure of classification performance, particularly when class distributions are imbalanced.
The empirical evaluation shows that our feature-engineered model yields predictions with substantial veracity. The models consistently achieved MCC scores greater than 0.6 and accuracy scores greater than 0.8 on the held-out test data, indicating strong predictive capability beyond random chance. The key contributions of this work do not end with the numbers, as the results democratize the forecasting; the training data is obtained from open OHLC data. Additionally, this research also prioritizes the transparency of the learned model. To this end, we employ Shapley values \cite{shapley}, a feature-explainability paradigm, to deconstruct the outputs of our models. This paradigm quantifies the contribution of each feature and helps us understand the utility of different genres of features employed in this research. 

The key contributions of this research are multifold. Firstly, to the best of our knowledge, this research is the first to model raw open-high-low-close (OHLC) signals to predict the open index on the next trading day. Secondly, the integration of the Shapley paradigm metamorphoses the procedure from a black box into a transparent tool for understanding the drivers of the healthcare indices. Finally, the research renders societal impact. It democratizes forecasting: the entire dataset is derived from publicly available OHLC data, extending sophisticated analysis to the general public (and beyond the learned stakeholders).  

The rest of this paper is organized as follows. In the next section, we describe the related works. In Section 3, we elaborate on the methodology for data sculpting. Sections 4 elaborates on the data source and computational methodology. In Sections 5 and 6, we present the experimental results and their discussion, respectively. We conclude the paper in Section 7.






\section{Related Works}

Stock markets form the backbone of modern economies \citep{stock-economy1, stock-economy2}, serving as a critical indicator of a nation's financial health. For decades, researchers have been intrigued by the interplay of factors that drive market fluctuations \citep{stock1,review-stock}, which is akin to a puzzle lying at the intersection of data and human behavior \citep{investor-psychology}. The early days of investigation encompassed aspects like macroeconomic fundamentals and the exploration of investor psychology and market sentiment \citep{stock-sentiment1,stock-sentiment2,stock-sentiment3,stock-sentiment4}. \citep{stock-sentiment1} explored the herding mentality of investors, which is deemed a significant cause of stock volatility, leptokurtosis, and non-IIDness. \citep{stock-sentiment2} investigated the association between investors' sentiment and stock returns and found a negative correlation. Their study suggests that a high confidence rendered a lower return in the future. \citep{stock-sentiment3} built their research narrative around investors' sentiment in a top-down approach, focusing on aggregate sentiment and tracing its effects on the market and individual stocks. Their research reveals a differential effect of sentiment on different stocks: low-capitalization, unprofitable, highly volatile, and young stocks are more vulnerable to investor sentiment. The findings of \citep{stock-sentiment4} are in congruence with the former, revealing that investor sentiment shows a stronger effect on securities whose valuations are highly subjective and difficult to arbitrage.

More recently, a paradigm shift has been observed, with researchers increasingly turning to sophisticated quantitative and statistical tools to decode market behavior \citep{stock-quant1,stock-quant2}. This movement, coupled with advances in computational power and the availability of financial datasets, shifted the focus from qualitative theories to the identification of empirical patterns hidden in the noise \citep{stock-dss1}. \citep{statistical-stock1} investigated three liquid stocks of the Paris Bourse. The study showed that incoming limit order prices follow a scaling behavior around the current price, with a diverging mean. This reflects market participants' belief that very large price variations can occur within a relatively short period. \citep{statistical-stock2} focused on comparing the effectiveness of independent indicators — such as price-to-sales, market-to-book, earnings per share, dividend yield, and market capitalization — against the dependent variable of stock returns. The study remained inconclusive regarding the degree of correlation of the factors with the stock returns, but it paved the way for understanding the coaction of the factors. The study in \citep{statistical-stock3} explored the applicability of the stochastic dominance statistic for diametrically opposite categories of investors -- risk avertors and risk takers, and analyzed the dominance relation in the US and Chinese markets. 

Recently, with the advent of computational tools, attention shifted to automating the learning process \citep{ml-decision2}. \citep{ml-decision} used long-term and short-term memory aspects of RNN for text mining on the organizations' announcements and their effect on the rise or fall of stock prices. \citep{ensemble-stocks} integrated empirical model decomposition and extreme learning machine to render an efficient model for stock prediction. \citep{long-term-prediction} used a nonparametric smoother with the covariates (the smoothing parameter selected by cross-validation) to forecast the long-term returns of stocks. A tensor-based framework is introduced in \citep{tensor-stock}, where the goal is to capture the individual associations between multiple information sources and the rise and fall of stock prices. \citep{voice-trading} explored the voice calls of managers and used a two-stage deep learning model to reduce the forecast errors, thereby yielding economic advantages in options trading. \citep{esg-stock} is research along the same line, which incorporates Environmental, Social, and Governance (ESG) sentiment and technical indicators to predict stock market behavior. \citep{self-adaptive} has proposed a self-adaptive trading strategy that is based on ML-based stock prediction. Despite the high volume of research on the aggregated market, its practical utility has often been limited. Consequently, scholars have shifted to sector-specific analysis.

The stock market aggregates stocks from different sectors. The phenomenon of volatility change is not uniform and depends on the sectoral composition. Consequently, a substantial segment of the literature is devoted to unraveling and understanding this variance. Several studies have investigated the sectoral volatility of nation and region-specific exploration. \citep{vol-gulf} studied the volatility fluctuations of equity sectors of the Gulf market, while \citep{vol-north-america} explored the impact of geopolitical events on sector-specific volatility spillovers in the US market. The study of volatility and its spillover leads to the demarcation of cyclical and defensive sectors. The defensive sectors are driven by consistent, day-to-day needs like consumer staples and healthcare \footnote{\url{https://www.alliancebernstein.com/corporate/en/insights/investment-insights/healthcare-stocks-an-innovative-antidote-for-volatile-times.html}}. Their low $\beta$-value (measures its volatility and systematic risk compared to the overall market) manifests a lower level of fluctuations and weaker co-movement with the whole market \citep{beta-healthcare1}.
In recent years, several studies have examined the factors influencing the movement of healthcare and pharmaceutical stocks \citep{pharma-trend1}. However, the current literature remains limited in scope: i] It examines only a few individual stocks without considering sectoral dynamics, and ii] It lacks an integrated understanding of the factors influencing sectoral index movements. To bridge this gap, our research aims to understand the role of different types of features related to the index movement of the healthcare sector and to identify those that correlate strongly with it.

\section{Methodology for data sculpting}

\subsection{Motivation}

The accurate prediction of the upcoming movement of the healthcare index through the knowledge of historical data can help investors make informed financial decisions. In this research, we extract information from prior data (indices) and use it to build a model to predict the upcoming trend (increase or decrease) in the open indices. For this task, we consider several feature genres. In our data, each row corresponds to a working day with the market being open, and each column represents a distinct feature used in the task. The upcoming subsections describe how the class label and the features are constructed. 
\subsection{Data labeling}

In this work, we predict the rise/ fall of the upcoming trading day's open index of the healthcare sector with respect to its previous day's index. Note that there are four sets of indices, and we can obtain four distinct class labels by comparing the indices of a day with the opening index of the next day. Hence, we generate four different class labels --${{y}}_\text {op}^\text {op}(t)$, ${{y}}_\text {hi}^\text {op}(t)$, ${{y}}_\text {lo}^\text {op}(t)$, and ${y}_\text{cl}^\text {op}(t)$ by comparing the open index of Day $(t+1)$ with the open ($op$), high ($hi$), low ($lo$) and close ($cl$) indices of Day $(t)$ respectively. Let $\text {P}_\text {op}(t)$ denote the open index of Day $(t)$. We will train a dedicated classifier for each class label (elaborated in Section \ref{subsec:method}). The generic formula for obtaining ${\text {Class}}_\text {op}^\text {op}(t)$ is given below.

\begin{equation}
    {{y}}_\text {op}^\text {op}(t) = 
    \begin{cases}
        1 & \text {if } \text {P}_\text {op}(t+1) > \text {P}_\text {op}(t) \\
        0 & \text{otherwise}
    \end{cases}
\end{equation}

Similarly, we can compute the class labels ${{y}}_\textrm{hi}^\textrm{op}(t)$, ${{y}}_{\textrm{lo}}^{\textrm{op}}(t)$, and ${{y}}_\textrm{cl}^\textrm{op}(t)$ by comparing the $\text {P}_\text {op}(t+1)$ with $\text {P}_\text {hi}(t)$, $\text {P}_\text {lo}(t)$ and $\text {P}_\text {cl}(t)$, respectively. 

\textbf{Illustration}: To generate the class label of Day 1 for open-high difference, denoted by ${{y}}_\textrm{hi}^\textrm{op}(1)$, we compare $\text {P}_\text {op}(2)$ with $\text {P}_\text {hi}(1)$: if $\text {P}_\text {op}(2) > \text {P}_\text {hi}(1)$,  class label is set to 1, and 0, otherwise.

Note that, while predicting for Day $(t+1)$, we will not know the indices of Day $(t+1)$ (as they are forthcoming). The task will be to predict if the class label is 1 (if the open index of Day $(t+1)$ will exceed the previous day's index) or 0 (if the open index of Day $(t+1)$ will be less than the previous day's index) based on the previous day's (Day $t$) indices. This data labeling converts the problem of forecasting into a supervised learning one.  
\\ \emph{Significance}: The model's significance lies in its predictive timing: it can forecast the direction of the next day's opening price ${{y}}_\text {op}^\text {*}(t))$ as soon as the current trading day closes. It is because the features needed for the next day's prediction are finalized at the market closing of the current day. The prior information about the market's movement can help the stakeholders make their investment decisions.

\subsection{Intrinsic features}

The intrinsic features for a day are obtained from the immediately previous day's open, high, low, and close indices. The indices of Day $t$ serve as the intrinsic feature for Day $t$. With four distinct index types present, we obtain four intrinsic features. 
Let the intrinsic feature set of Day $t$ be denoted by $\mathbf{x}_\text{int}(t)$.

\begin{equation}
\mathbf{x}_\text{int}(t) = [\textrm{P}_\textrm{op}(t), \textrm{P}_\textrm{hi}(t), \textrm{P}_\textrm{lo}(t), \textrm{P}_\textrm{cl}(t)]^\top \quad 
\end{equation}

\textbf{Illustration}: If we consider July 19, 2024 (working day) as Day $t$, the indices of July 19, 2024, constitute the intrinsic features of Day $t$. The task is to predict if the open index will rise or fall at Day $(t+1)$. The intrinsic features offer a snapshot of the immediately available market scenario in its crude form. They can offer important facets into the upcoming market scenario, but are usually suboptimal in elucidating the gross market behavior. 

\subsection{Volatility}

Volatility in markets is perceived to be a crucial determinant of the index movements by traders \citep{volatility-trading1,volatility-trading2}. To this end, we include the volatility of the healthcare market as a feature in our training data. To render a human-interpretable measure of volatility, we compute features using \textit{Donchian Channel (\cite{donchian}), Bollinger Bands (\cite{bollinger}), and Keltner Channel (\cite{keltner})}, which are standard frameworks in market analysis.

Traders rely on these parameters to assess volatility and trading scenarios, enabling them to make informed financial decisions \citep{volatility-metrics}. In this research, we calculate these parameters for each trading day and incorporate them as features for our own model-building exercise.

\begin{itemize}

\item \textbf{Donchian Channel}:

The Donchian Channel \footnote{\url{https://www.tradingview.com/chart/IBKR/rO4ruhxQ-Donchian-Channel-Strategy-like-The-Turtles-Traders/}}, developed in the 1960s, is a technical indicator that helps traders understand market volatility and identify breakout and reversal points. Technically, it provides three bands: $\text{Upper}_{\text{DC}}(t)$ -- represents the highest high over a specified look-back period (we have taken 20 periods), $\text{Lower}_{\text{DC}}(t)$ -- the lower band captures the lowest low over the same period, and $\text{Middle}_{\text{DC}}(t)$, the middle band is calculated as the average of the upper and lower bands. The Donchian Channel encapsulates the index movement within this channel. The bands widen during phases of high volatility and contract during phases of low volatility. 

\begin{equation}
\begin{aligned}
&\text{Upper}_{\text{DC}}(t) = \max(\textrm{P}_\textrm{hi}(t-n), \textrm{P}_\textrm{hi}(t-n+1), \ldots, \textrm{P}_\textrm{hi}(t)) \quad , \\[6pt]
&\text{Lower}_{\text{DC}}(t) = \min(\textrm{P}_\textrm{lo}(t-n), \textrm{P}_\textrm{lo}(t-n+1), \ldots, \textrm{P}_\textrm{lo}(t)) \quad , \\[6pt]
&\text{Middle}_{\text{DC}}(t) = \frac{\text{Upper}_{\text{DC}}(t) + \text{Lower}_{\text{DC}}(t)}{2} \quad 
\end{aligned}
\end{equation}

Let $\mathbf{x}_\text{DC}(t)$ be the band information with respect to the Donchian Channel at Day $t$.

\begin{equation}
\mathbf{x}_\text{DC}(t) = [\text{Upper}_{\text{DC}}(t),\text{Lower}_{\text{DC}}(t),\text{Middle}_{\text{DC}}(t)]^\top \quad 
\end{equation}
 

\item \textbf{Bollinger Bands}:
It is another volatility-capturing tool adopted by traders to assess market conditions, potential breakout opportunities, and overbought or oversold signals \footnote{\url{https://www.tradingview.com/chart/US500/IavHLFpW-Using-Bollinger-Bands-to-Gauge-Market-Trends-and-Volatility/}}. Similar to the Donchian Channel, there are three bands -- the middle band is derived from the 20-period simple moving average (SMA), while the upper and lower bands are placed at a set number of standard deviations (usually two) above and below the SMA. The lower and upper bands diverge during high volatility and contract during low volatility. This dynamic movement enables the traders to identify compressions (Bollinger Band squeezes) that often precede significant index breakouts. 
Simple Moving Average (SMA) computes the arithmetic mean of indices over $n$ periods at Day $i$: 
\begin{equation}
\text{SMA}_n(t) = \frac{1}{n}\sum_{i=t-n+1}^t \text{P}_\text{cl}(i)
\end{equation}

\begin{equation}
\begin{split}
\text{Middle}_{\text{BB}}(t) &= \text{SMA}_n(t)\footnotemark, \\
\sigma_n(t) &=
\sqrt{\tfrac{1}{n}
\sum_{i=t-n+1}^t
\bigl(\text{P}_{\text{cl}}(i)-\text{SMA}_n(i)\bigr)^2}, \\
\text{Upper}_{\text{BB}}(t) &= \text{SMA}_n(t) + k\,\sigma_n(t), \\
\text{Lower}_{\text{BB}}(t) &= \text{SMA}_n(t) - k\,\sigma_n(t).
\end{split}
\end{equation}
\\, where $k$ is the weight, usually chosen as 2, or between 1 and 3.

Let $\mathbf{x}_\text{BB}(t)$ be the band information with respect to Bollinger Bands at Day $t$.
\begin{equation}
\mathbf{x}_\text{BB}(t) = [\text{Upper}_{\text{BB}}(t),\text{Lower}_{\text{BB}}(t),\text{Middle}_{\text{BB}}(t)]^\top \quad 
\end{equation}


\item \textbf{Keltner Channel}:
It draws an envelope around the moving average of indices, considering a prefixed period (we have taken a 20-day period)\footnote{\url{https://www.ig.com/en/trading-strategies/how-to-trade-using-the-keltner-channel-indicator-191119}}. The exponential moving average (EMA) of the prefixed period serves as the middle line for a day, while the upper and lower bands are placed at a distance proportional to the average true range (ATR) over the same period.

Exponential Moving Average: 
\begin{equation}
\text{EMA}_n(t) = \alpha \text{P}_\text{cl}(t) + (1-\alpha)\text{EMA}_{t-1}\textrm{, where }\alpha=\frac{2}{n+1}
\end{equation}

\begin{equation}
\begin{split}
\text{Middle}_{\text{KC}}(t) &= \text{EMA}_n(t)\footnotemark, \\ 
\text{Upper}_{\text{KC}}(t) &= \text{Middle}_{\text{KC}}(t) + k \cdot \text{ATR}_n(t), \\ 
\text{Lower}_{\text{KC}}(t) &= \text{Middle}_{\text{KC}}(t) - k \cdot \text{ATR}_n(t), \\
\text{ATR}_n(t) &= \frac{1}{n}\sum_{i=t-n+1}^t \text{TR}(i), \\
\text{TR}(t) &= \max\Bigl(
\text{P}_{\text{hi}}(t) - \text{P}_{\text{lo}}(t), \\
&\quad \left|\text{P}_{\text{hi}}(t) - \text{P}_{\text{cl}}(t-1)\right|, \\
&\quad \left|\text{P}_{\text{lo}}(t) - \text{P}_{\text{cl}}(t-1)\right|
\Bigr).
\end{split}
\end{equation}

Let $\mathbf{x}_\text{KC}(t)$ be the band information for the Keltner Channel at Day $t$.

\begin{equation}
\mathbf{x}_\text{KC}(t) = [\text{Upper}_{\text{KC}}(t),\text{Lower}_{\text{KC}}(t),\text{Middle}_{\text{KC}}(t)]^\top \quad 
\end{equation}

The band information helps in tracking index movements, identifying support/resistance levels, and also signals breakouts when indices move beyond the band levels.

\end{itemize}

The three tools considered in this work are motivated to capture the volatility in the index movements (and also the historical index movements). Despite this underlying similarity between the three, they provide distinct market insights, each indicative of different trading scenarios. The Donchian channel, the most interpretable among the three, efficiently identifies breakout levels in the market, but is prone to false breakouts (when the low or high index shows sudden changes). Information derived from Bollinger Bands and Keltner Channel is immune to this disadvantage as they rely on moving averages; the contribution of one or a few upward or downward index spikes is dampened by averaging. Bollinger Bands are effective in predicting a surge in indices after a stagnant market scenario; however, they are prone to generating false overbought (or underbought) signals under strong directional momentum. Lastly, by virtue of the exponential function, Keltner Channel is more immune than Bollinger Bands to raising false overbought (or underbought) signals, but is less sensitive to sudden volatility spikes. To have their interpretable but diverse insights in the model, we include bands' information from all three in the feature set. 
Let $\mathbf{x}_\text{hist}(t)$ be the historical information about volatility at Day $t$. It consists of features extracted from all three bands.

\begin{equation}
\mathbf{x}_\text{hist}(t) = [\mathbf{x}_\text{DC}(t),\mathbf{x}_\text{BB}(t), \mathbf{x}_\text{KC}(t)]^\top \quad 
\end{equation}

\subsection{Nowcasting Methods}
A glimpse of the current market scenario is provided by the intrinsic features, $\mathbf{x}_\text{int}(\dot)$ (as discussed in Section 3.3). On the other hand, the features related to historical prices summarization (as discussed in the previous section), $\mathbf{x}_\text{hist}(\dot)$, give a comprehensive picture of the market's movement over a past period. They offer a deep understanding of the volatility, stationarity, and movement of the indices. It is important to note that only four distinct indices are known for any trading day, including the current trading session, but the movement in the current day cannot be explicitly inferred from these four data points. Consequently, both categories of features are suboptimal for providing finer details of intraday market momentum on the current trading day.  
To address this gap and unfold the intraday narrative, we include the logarithm of the index ratios (with respect to open indices) in the feature set. They are $ln{(\frac{P_\textrm{hi}}{P_\textrm{op}})}$, $ln{(\frac{P_\textrm{lo}}{P_\textrm{op}})}$ and $ln{(\frac{P_\textrm{cl}}{P_\textrm{op}})}$.

\begin{subequations}
    \begin{align}
        r_{op}^{hi}(t) &= ln{\left(\frac{{P}_\textrm{hi}(t)}{{P}_\textrm{op}(t)}\right)}, \label{eq:1} \\
        r_{op}^{lo}(t) &= ln{\left(\frac{P_\textrm{lo}(t)}{P_\textrm{op}(t)}\right)}, \label{eq:2} \\
        r_{op}^{cl}(t) &= ln{\left(\frac{P_\textrm{cl}(t)}{P_\textrm{op}(t)}\right)}. \label{eq:3}
    \end{align}
\end{subequations}
The three values $r_{op}^{hi}(t), r_{op}^{lo}(t)$, and $r_{op}^{cl}(t)$ serve as nowcasting tools for predicting the rise-or-fall of index on the upcoming trading day. The combinations of their values can provide an array of information about movement in the current trading day's market, thereby contributing to the robustness of the model. Some of those perceptions are described as follows.

\begin{itemize}
    \item $r_{op}^{cl}(t)>0$ shows that the market closed at a value higher than it opened, $r_{op}^{cl}(t)<0$ indicates the opposite.
     \item $r_{op}^{lo}(t)\sim0$ yields that the market remained higher than the opening throughout the day.
     \item $r_{op}^{hi}(t)\sim0$ indicates that the market remained lower than the opening throughout the day.
    \item $r_{op}^{lo}(t)<0$ or $r_{op}^{hi}(t)>0$ (by some substantial margin) indicates a considerable movement in the market, where the low index fell below the open index and the high index rose above the open index.  
    \item $|r_{op}^{hi}(t)|\sim0, |r_{op}^{lo}(t)|\sim 0$, and $|r_{op}^{cl}(t)|\sim 0$ indicates a stationary market, where all four indices remain within a small range, thereby, rendering a close to 0 value of their logarithm ratios. 
\end{itemize}

To imbibe these reasoning into our training model, we include them to form a \emph{nowcasting} feature set, $\mathbf{x}_\text{now}(t)$. From the ML perspective, $\mathbf{x}_\text{now}(t)$ helps the model focus on the intraday movement and momentum of indices on the current trading day. 

\begin{equation}
\mathbf{x}_\text{now}(t) = [r_{op}^{hi}(t),r_{op}^{lo}(t),r_{op}^{cl}(t)]^\top \quad 
\end{equation} 

The nowcasting features capture the intraday price movements in the immediately previous trading day. This movement has the potential to influence and bridge the price movements in the upcoming trading day, and accordingly, we explore it as a feature for a price movement indicator.
\subsection{Final set of features}
The final set of features is formed by taking the concatenation of the intrinsic features ($\mathbf{x}_\text{int}$), features derived from market's volatility ($\mathbf{x}_\text{vol}$), and nowcasting tools ($\mathbf{x}_\text{now}$). 
\begin{equation}
\mathbf{x}(t) = \begin{bmatrix} \mathbf{x}_\text{int}(t) \\ \mathbf{x}_\text{hist}(t) \\ \mathbf{x}_\text{now}(t) \end{bmatrix}
\end{equation}
$\mathbf{x}(t)$ is used in building the final classifier. The final set of features encompasses diverse aspects: intrinsic prices of the immediately previous trading day, intraday price movements on the same day, and the historical features that capture the volatility in the twenty-day window. We explore whether this rich feature set can effectively associate with the target class of rise/ fall of prices and predict it for the upcoming trading day.

\section{Data Source and Computational Methodology}
\subsection{Data Collection}
This research focuses on predicting the opening trend of healthcare indices. To study the utility of our method, we include two datasets, which are derived from actual markets -- i] Healthcare index from Standard and Poor's (S$\&$P) 500, United States of America, the largest economy of the world (representative of a mature market), and, ii]  Healthcare index from Standard and Poor's (S$\&$P) Bombay Stock Exchange (BSE), India, representative of the emerging economies. S$\&$P 500 is a stock market index which tracks the stock performance of 500 leading companies listed on stock exchanges in the United States; the market capitalization of the enlisting reached US $\$$ 55 trillion in July 2025 \footnote{\url{https://thetradable.com/stocks/sp-500-market-cap-hits-record-557-trillion-historic-rally-continues-ig}}. S$\&$P BSE Sensex is an Indian free-float, market-weighted stock market index of 30 stable and financially robust companies listed on the BSE. The total market capitalization of listings in BSE is US $\$$ 5.6 trillion as of September 2024 \footnote{\url{https://www.ceicdata.com/en/indicator/india/market-capitalization}}.  

For both countries, we have considered five fiscal years, from 1st April 2019 to 31 March 2024. In this period, S$\&$P 500 (U.S.A.) recorded 1256 trading days and the same figure was 1237 for BSE (India). Local events and market closures are accountable for the minor variations in trading days. For each trading day $t$, their open, high, low, and close indices have served as the intrinsic features. The calculation of historical features requires a window of 20 days, out of which 19 days are before the day of calculation. Hence, we have to exclude the initial 19 days from our calculation. Day $t$'s class label is derived by comparing its index (open/ high/ low/ close) with the open index of the next trading day $(t+1)$. As said, we cannot compute the class label for the last day of the period (as it depends on the next trading day), so we exclude it. So, a total of 20 trading days is excluded from each dataset. Hence, for S$\&$P 500 (U.S.A.) and BSE (India), our datasets consist of 1236 and 1217 trading days, respectively. 
The basic description can be found in Table \ref{tab:data}.

\begin{table*}[h]
\centering
\caption{Summary of Datasets}
\label{tab:data}
\begin{tabular}{lcc}
\toprule
\multirow{2}{*}{} & \multicolumn{2}{c}{Healthcare index}\\
\cmidrule(lr){2-3}
 & S$\&$P 500, U.S.A. & BSE, India \\
\midrule
Period & \multicolumn{2}{c}{Apr 1, 2019 -- Mar 31, 2024} \\
No. of trading days & 1256 & 1237 \\
No. of days discarded & \multicolumn{2}{c}{------20------} \\
No. of points in the dataset & 1236 &1217 \\
No. of training points & 989 & 974\\
No. of test points & 247 & 234\\
\bottomrule
\end{tabular}
\end{table*}

We consider each dataset as a time series, as observations consist of indices, which are ordered chronologically. The first 80$\%$ days are considered for training the model, and the remaining 20$\%$ are used to test the model. This allows the model to learn the temporal associations. We have incorporated one-step-ahead rolling window approach for prediction instead of multi-step-ahead forecasting (\citep{rolling-window}).  We evaluate the model's predictive capability by calculating the accuracy and MCC scores on predictions. 

\subsection{Method}
\label{subsec:method}
In this work, we build ML-based models to predict the rise/ fall of open index of the healthcare index in the upcoming trading day. The datasets for our model are derived from S$\&$P 500, U.S.A., and BSE, India. While constructing the feature set, we use the intrinsic indices as well as some derived features, as described in the previous section. Let $\mathcal{D}^{*}$ be the dataset, consisting of the points (represented by their features), $\mathbf{x}(\cdot)$, and the corresponding class labels $y^{op}_{*}(\cdot)$.

\begin{equation}
\mathcal{D}^{*} = \{ (\mathbf{x}(i), {y}^{op}_{*}(i) \}_{i=1}^N, \quad N = |\mathcal{D}^{*}|, * \in \{op, hi, lo, cl \}
\end{equation}

We split $\mathcal{D}^{*}$ into a training set, $\mathcal{D}^{*}_{tr}$ and a test set, $\mathcal{D}^{*}_{te}$. Points in $\mathcal{D}^{*}$ are arranged chronologically, its first $80\%$ points form $\mathcal{D}^{*}_{tr}$ and the remaining (latest) $20\%$ points form $\mathcal{D}^{*}_{te}$. Note that, depending on $*$, the dataset changes. The change happens due to the change in the target class only; the feature set remains the same across all cases.

In this work, we consider four different problem cases by comparing four types of indices of Day $t$ with the open index of Day $(t+1)$. They are -- i] comparison of the open indices of Day $(t+1)$ and Day $t$, ii] comparison of the open index of Day $(t+1)$ with the high index of Day $t$, iii] comparison of the open index of Day $(t+1)$ with the low index of Day $t$, and, iv] comparison of the open index of Day $(t+1)$ with the close index of Day $t$. For each of these subproblems, we train a different classifier model using its dedicated training dataset and obtain predictions on its dedicated test set.

We train model $\mathcal{M}^{*}$ to predict the increase or decrease of upcoming indices with respect to their corresponding previous day's $*$ indices, where $* \in \{op, hi, lo, cl\}$. We use $\mathcal{D}^{*}_{tr}$ to construct $\mathcal{M}^{*}$. As the cardinality of the set $\{op, hi, lo, cl\}$ is \(four\), we train \(four\) different models -- $\mathcal{M}^{op}$, $\mathcal{M}^{hi}$, $\mathcal{M}^{lo}$ and $\mathcal{M}^{cl}$. 

The model $M^{*}$ is trained on $\mathcal{D}^{*}_{tr}$ which consists of $\alpha$ points:
\begin{equation}
M^{*} = \mathcal{A}\left( \{(x(i), y_{*}^{\text{op}}(i))\}_{i=1}^\alpha \right),
\end{equation}
where \(\mathcal{A}\) is the learning algorithm, \(x(i)\) is the \(i\)-th input, \(y_{*}^{\text{op}}(i)\) is the corresponding class label of point \(x(i)\), \(\alpha\) denotes the total number of training points.

\subsection{Classifiers used}
We have employed an array of classifiers to assess the discriminative power of the crafted feature set. When selecting classifiers, we considered a mix of interpretable and black-box models. There are eight different classifier types – four with a substantial degree of interpretability (Decision Tree, Gaussian Naive Bayes, k-Nearest Neighbor, and Logistic Regression), and four with low interpretability (eXtreme Gradient Boosting (XGB), Multi Layer Perceptron (MLP), CatBoost, and ExtraTrees). The choice of classifiers is made to ensure that the prediction model is free of algorithmic bias, thereby revealing the true capabilities of the diverse feature genres.

We used Python implementations of all classifiers (\cite{sklearn-classifiers}). For the Decision Tree, we set $max_depth=10$ and $max_features=5$. Gaussian Naive Bayes is used at its default configuration. The neighborhood size in the $k$-nearest neighbor classifier is set to 5, and the \textit{liblinear} solver is employed in the Logistic Regression Classifier. We have used the default parameter configuration for the XGB Classifier. For the MLP Classifier, we have used 8 hidden layers, \textit{relu} activation function, and 1000 iterations. We set 1000 iterations and a learning rate of 0.1 for the CatBoost Classifier. For ExtraTreesClassifier, the criterion is set to \textit{cross-entropy} and the number of estimators is set to 1000. 

\subsection{Evaluation metrics}
We measure the veracity of predictions using accuracy scores and Matthews Correlation Coefficient (MCC) scores \cite{mcc1,mcc2}. Accuracy offers an intuitive, widely prevalent, but potentially misleading view of performance, as it can be inflated by market imbalances (for example, always predicting a rise in a bull market or a fall in a bearish market). Compared to accuracy, MCC is a far more robust metric, as it accounts for all classes of predictions. A high MCC score confirms the model's veracity in all scenarios, which is critical for the practical utility of the outcomes. The formula and reference levels for MCC scores are provided in \ref{sec:mcc}. For both metrics, higher scores indicate better performance. 
\section{Results}
\subsection{Performance of the classifiers}
In this section, we present the veracity of predictions on the rise or fall of the open index of the upcoming trading day. Like every ML model-building and evaluation process, this work has two principal steps: i] data sculpting from raw data on the open, high, low, and close indices, and ii] developing the underlying architecture of the classifier. An intrinsic data sculpting method is proposed in this work, incorporating three feature genres. We test their individual as well as cumulative efficiencies. A spectrum of different classifiers is explored to find out the one/s with the highest class discerning ability(ies). Figures 1-8 (a) and (b) record the performance of the methods. We present the results in a segregated fashion, with respect to i] index comparison (related to four different classes), ii] classifiers, and iii] features (related to).

\begin{itemize}
\item \textit{For which type of index-comparison, do we get the highest veracity in predictions?}
\\ This work explores four different tasks, prediction of the rise or fall of the upcoming trading day's open index with respect to the current day's i] open index, ii] high index, iii] low index, and iv] close index. 

Figures \ref{fig:acc-open} and \ref{fig:mcc-open} illustrate the accuracy and MCC scores for predicting the next day's open index movement—rise or fall—relative to the current day's \emph{open index}, respectively. In this case, the question addressed by the model is -- \emph{Will the market open at a higher level than it opened today?} The tabular data shows that the models can provide trustworthy predictions for both the markets (S$\&$P 500 and BSE). Many of the classifier architectures and feature combinations have yielded commendable accuracy and MCC scores, with accuracies of $0.80$ and MCCs of $0.60$. These data manifest the utility of the proposed framework (with the correct choice of features and classifiers) in predicting the rise or fall of the next day's open index (relative to the current day's open index) for the healthcare sector. 

Figures \ref{fig:acc-high} and \ref{fig:mcc-high} depict the performance of predicting the next day's open index movement—rise or fall—relative to the current day's \emph{high index}, respectively. In this case, the question addressed by the model is -- \emph{Will the market open at a higher level than the highest index seen today?}. The tables show that the prediction framework yields reliable results, with MCC and accuracy scores of $0.60$ and $0.80$, respectively, across several models. The healthcare index prediction for S$\&$P 500 has shown greater reliability than that of BSE, as the MCC scores for the former lie in the range $0.80$, whereas for the latter they lie in the range $0.65$.

Figures \ref{fig:acc-low} and \ref{fig:mcc-low} record the performance of predicting the next day's open index's rise or fall, relative to the current day's \emph{low index}. The implication of this task is whether the market will open higher than today's low. For this class of predictions, the accuracy ranges from $\tilde{0.90}$ for both markets, while the MCC scores remain in a moderately reliable range. For S$\&$P 500, the MCC scores range in $0.6$ for most of the well-doing classifiers, with only MLP peaking to $>0.8$. For BSE, performance is a notch lower, as the MCC scores remain around $\tilde{0.6}$ for the best-performing classifiers as well.

Lastly, the figures \ref{fig:acc-close} and \ref{fig:mcc-close} show the performance of predicting the next day's open index's rise or fall-relative to the current day's \emph{close index}. The implication of this prediction -- whether the market will open at a higher level tomorrow than today's closing price. Despite an admissible range of accuracy $(0.7-0.8)$ for a number of classifier-feature pairs (though lower than that of the previous three tasks), the MCC scores lie in the range $\tilde{0.0-0.1}$ for both the market (with only one exception in S$\&$P 500). 

Table \ref{tab:reliability-summary} summarizes the outcomes described above. This table shows the generic reliability of the proposed framework across the four subproblems described above. The table indicates whether an effective classifier existed for each subproblem, as evaluated by the two metrics. For accuracy and MCC scores, we set the reference levels to $0.8$ and $0.65$, respectively. The justification for the choices of values is elaborated in Sections \ref{sec:acc} and \ref{sec:mcc}.
The implications of the outcomes shown in the figures and Table \ref{tab:reliability-summary} are discussed in detail in Section 6.

\begin{figure*}[h!]
    \centering
    \begin{subfigure}[b]{\textwidth}
        \includegraphics[width=0.9\textwidth]{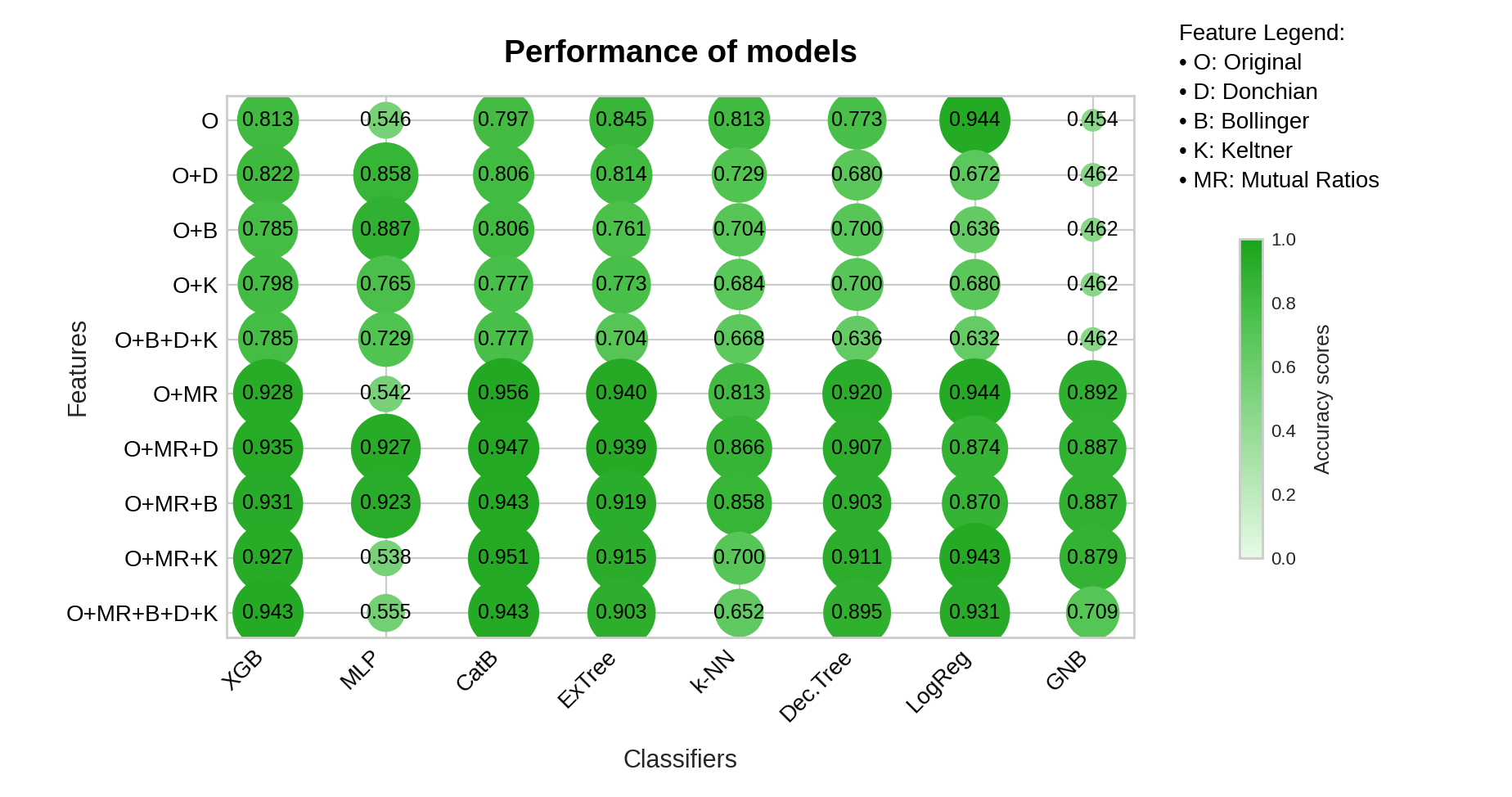}
        \caption{S$\&$P 500}
    \end{subfigure}
    \hfill
    \begin{subfigure}[b]{\textwidth}
        \includegraphics[width=0.9\textwidth]{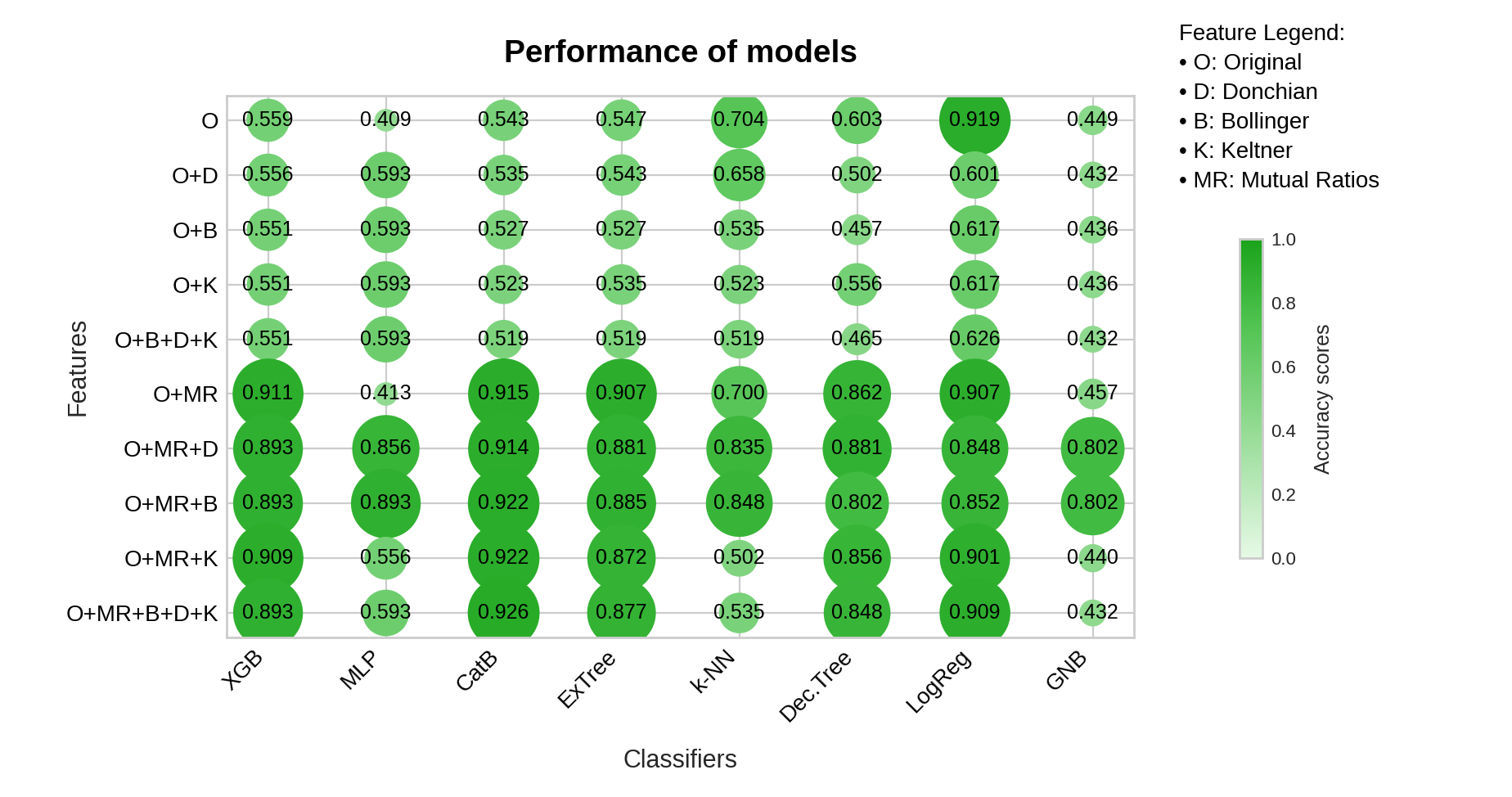}
        \caption{BSE}
    \end{subfigure}
    \caption{The plots present the accuracy scores for the \emph{open-open} across a feature-classifiers grid. Plots (a) and (b) are dedicated to S$\&$P 500 and BSE, respectively. The vertical axis shows different combinations of features, while the horizontal axis lists the classifiers. The higher the accuracy score, the better the performance. Consequently, larger circle radii and darker colors indicate better performance.}
    \label{fig:acc-open}
\end{figure*}

\begin{figure*}[h!]
    \centering
    \begin{subfigure}[b]{\textwidth}
        \includegraphics[width=0.9\textwidth]{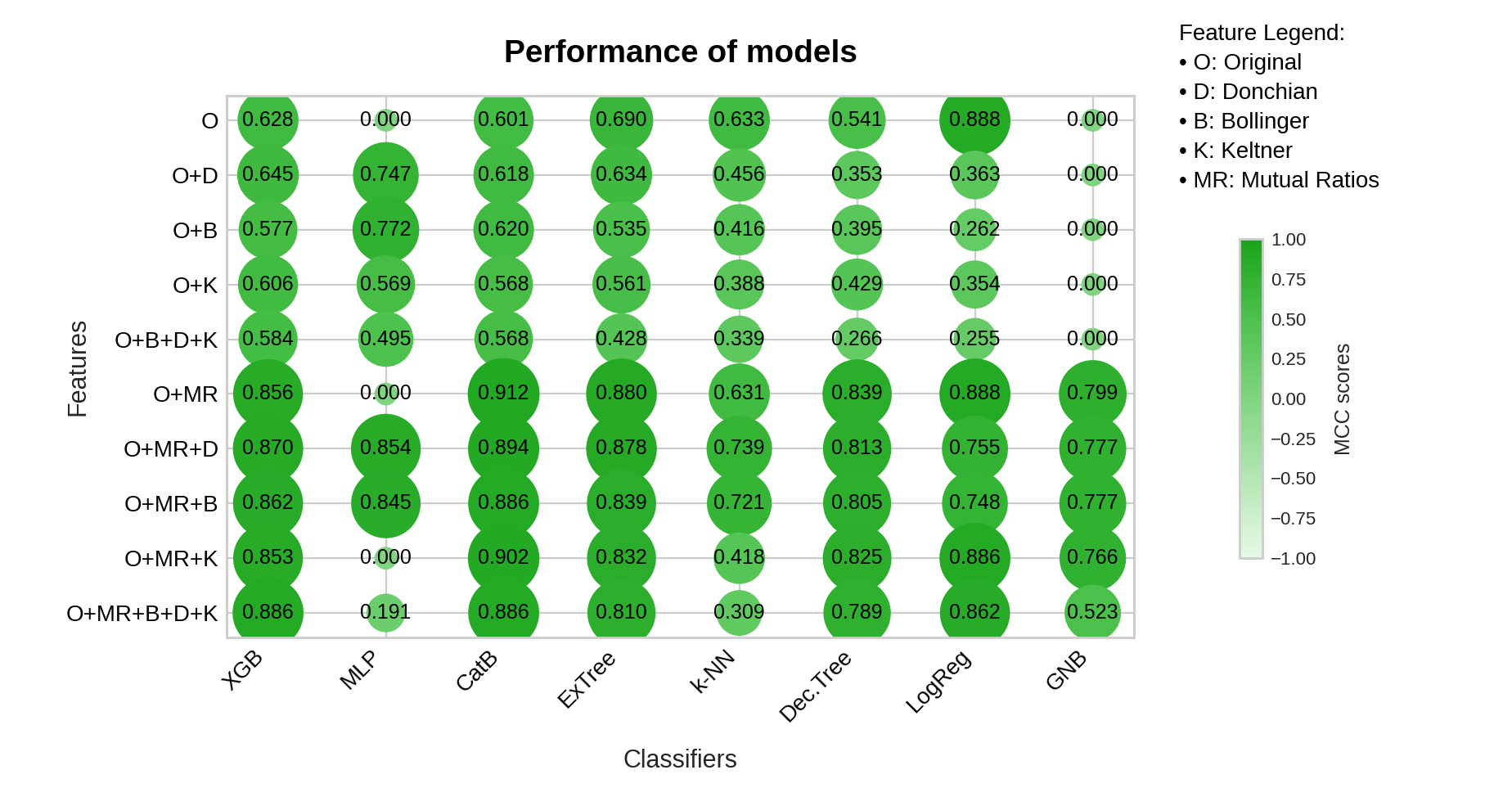}
        \caption{S$\&$P 500}
        \label{fig:fig1}
    \end{subfigure}
    \hfill
    \begin{subfigure}[b]{\textwidth}
        \includegraphics[width=0.9\textwidth]{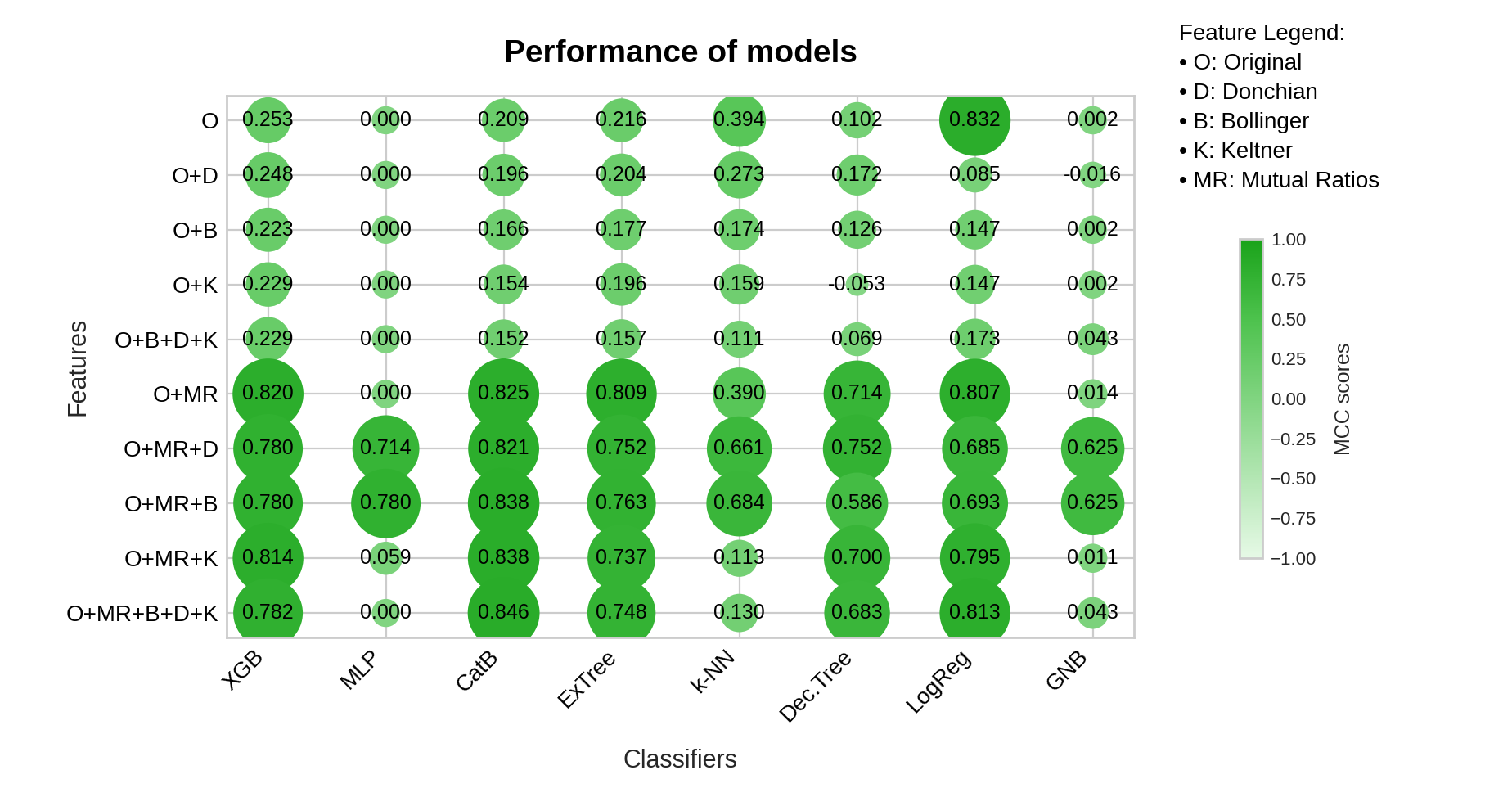}
        \caption{BSE}
        \label{fig:fig2}
    \end{subfigure}
    \caption{The plots present the MCC scores for the \emph{open-open} across a feature-classifiers grid. Plots (a) and (b) are dedicated to S$\&$P 500 and BSE, respectively. The vertical axis shows different combinations of features, while the horizontal axis lists the classifiers. The higher the MCC score, the better the performance. Consequently, larger circle radii and darker colors indicate better performance.}
    \label{fig:mcc-open}
\end{figure*}

\begin{figure*}[h!]
    \centering
    \begin{subfigure}[b]{\textwidth}
        \includegraphics[width=0.9\textwidth]{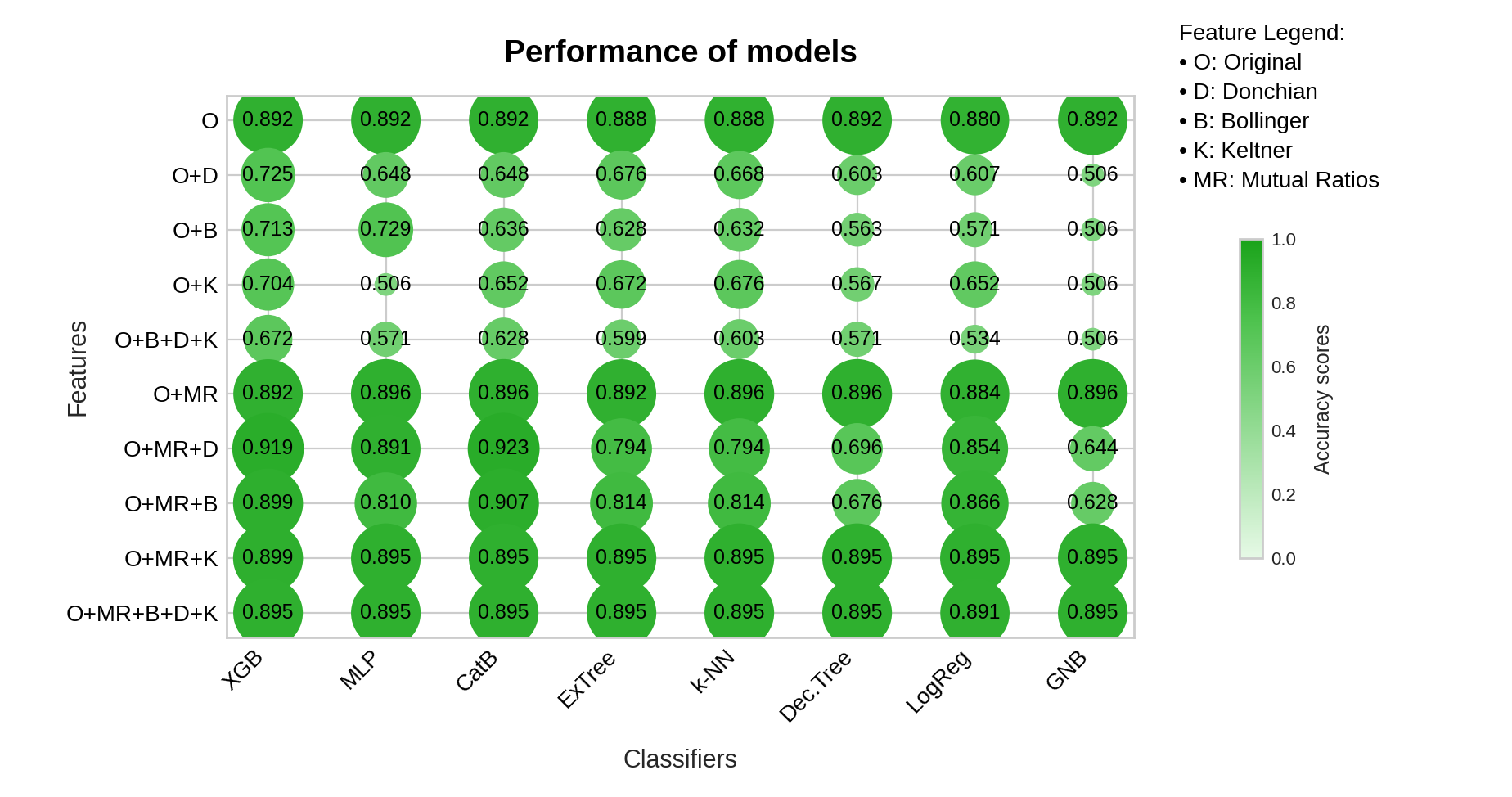}
        \caption{S$\&$P 500}
        \label{fig:fig1}
    \end{subfigure}
    \hfill
    \begin{subfigure}[b]{\textwidth}
        \includegraphics[width=0.9\textwidth]{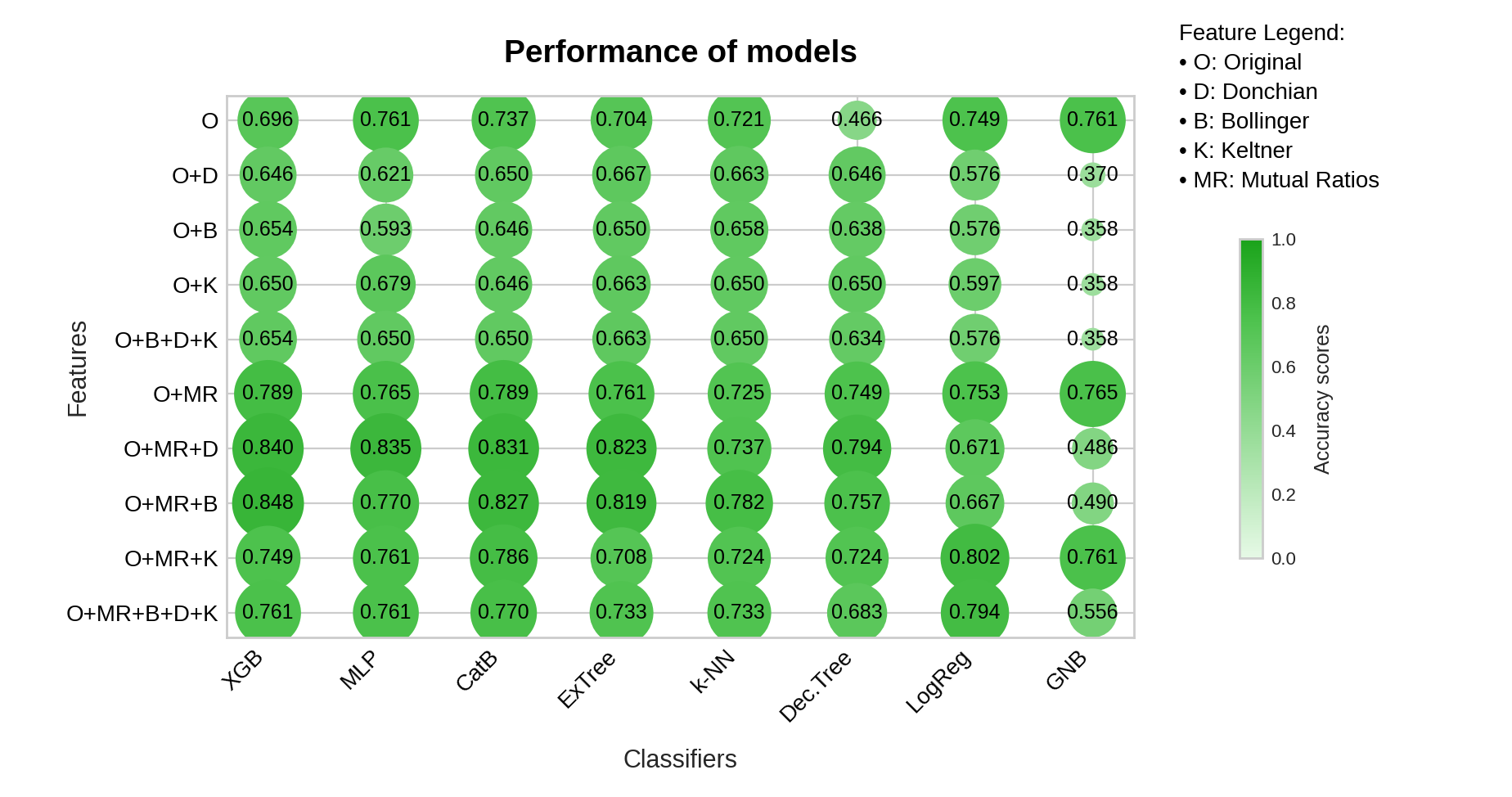}
        \caption{BSE}
        \label{fig:fig2}
    \end{subfigure}
    \caption{The plots present the accuracy scores for the \emph{open-high} across a feature-classifiers grid. Plots (a) and (b) are dedicated to S$\&$P 500 and BSE, respectively. The vertical axis shows different combinations of features, while the horizontal axis lists the classifiers. The higher the accuracy score, the better the performance. Consequently, larger circle radii and darker colors indicate better performance.}
    \label{fig:acc-high}
\end{figure*}

\begin{figure*}[h!]
    \centering
    \begin{subfigure}[b]{\textwidth}
        \includegraphics[width=0.9\textwidth]{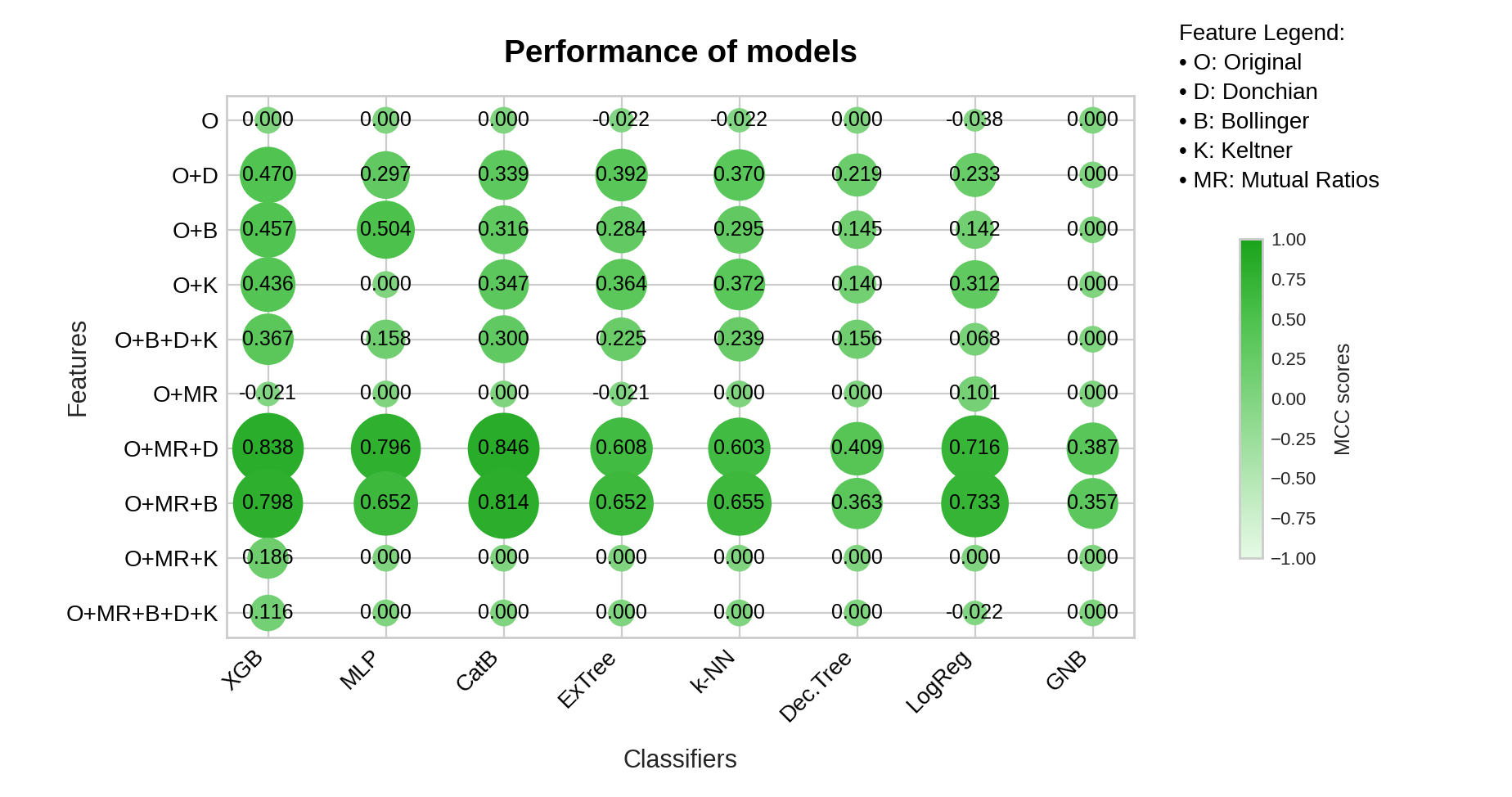}
        \caption{S$\&$P 500}
    \end{subfigure}
    \hfill
    \begin{subfigure}[b]{\textwidth}
        \includegraphics[width=0.9\textwidth]{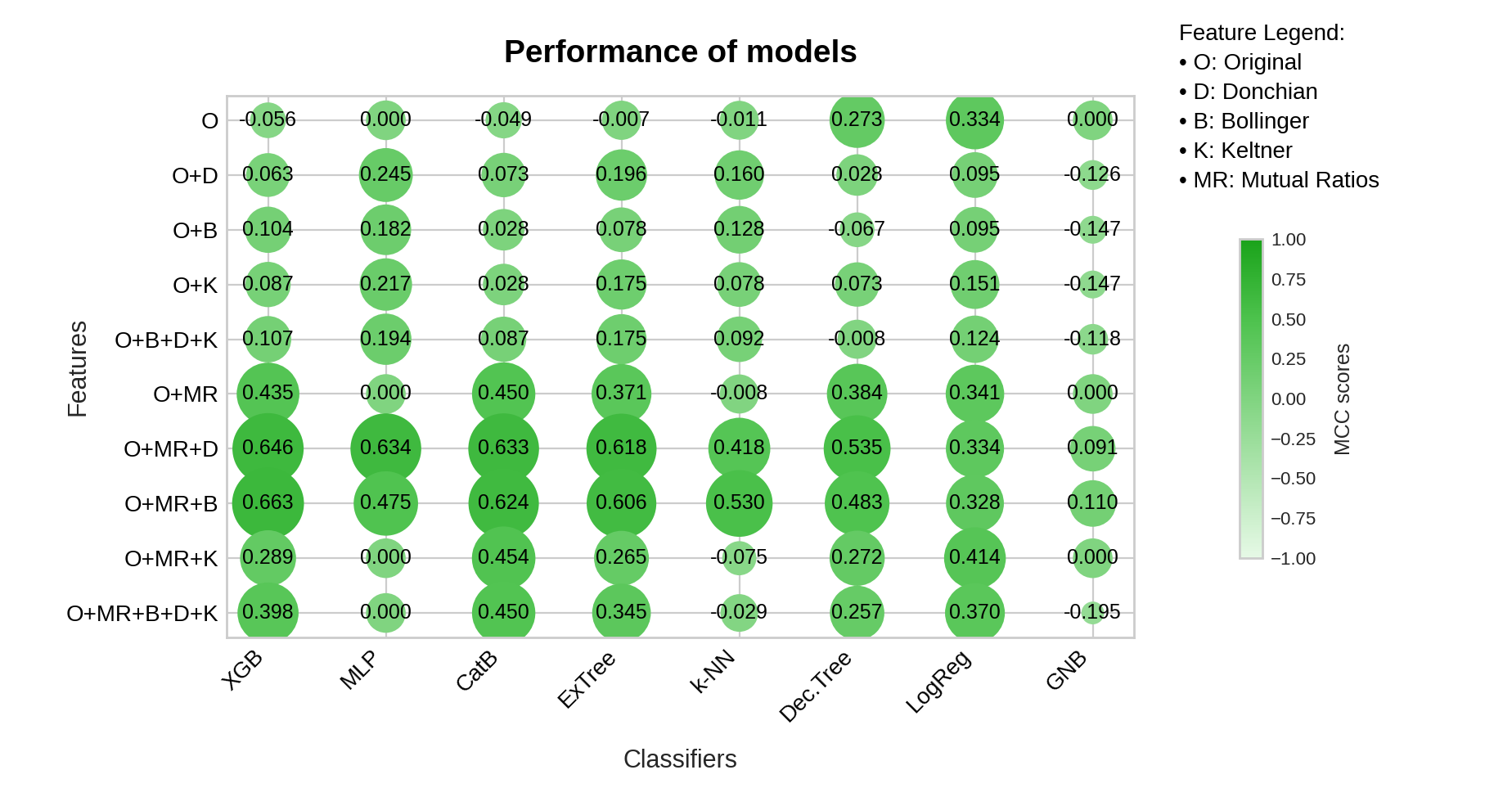}
        \caption{BSE}
    \end{subfigure}
    \caption{The plots present the MCC scores for the \emph{open-high} across a feature-classifiers grid. Plots (a) and (b) are dedicated to S$\&$P 500 and BSE, respectively. The vertical axis shows different combinations of features, while the horizontal axis lists the classifiers. The higher the MCC score, the better the performance. Consequently, larger circle radii and darker colors indicate better performance.}
    \label{fig:mcc-high}
\end{figure*}

\begin{figure*}[h!]
    \centering
    \begin{subfigure}[b]{\textwidth}
        \includegraphics[width=0.9\textwidth]{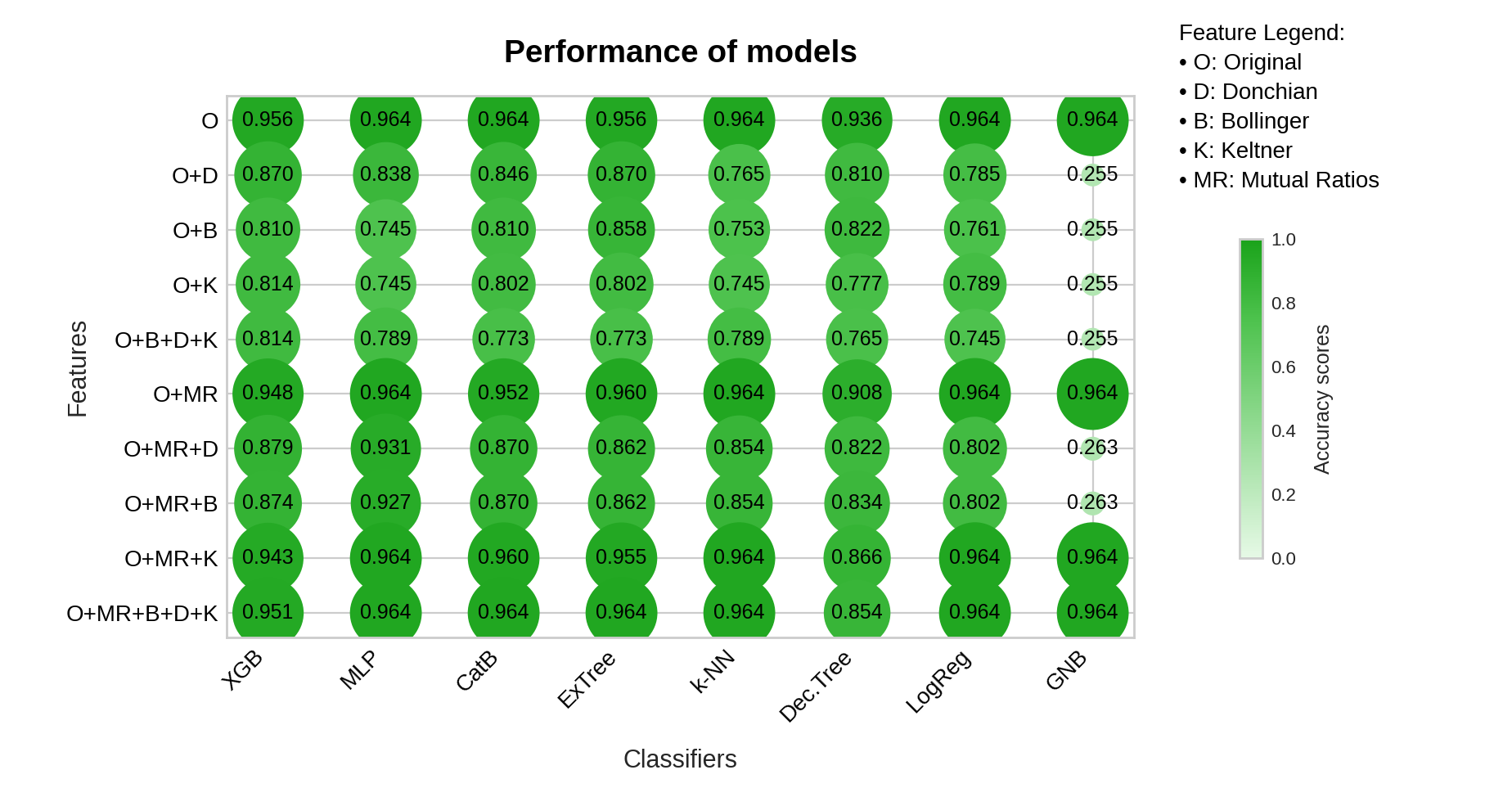}
        \caption{S$\&$P 500}
    \end{subfigure}
    \hfill
    \begin{subfigure}[b]{\textwidth}
        \includegraphics[width=0.9\textwidth]{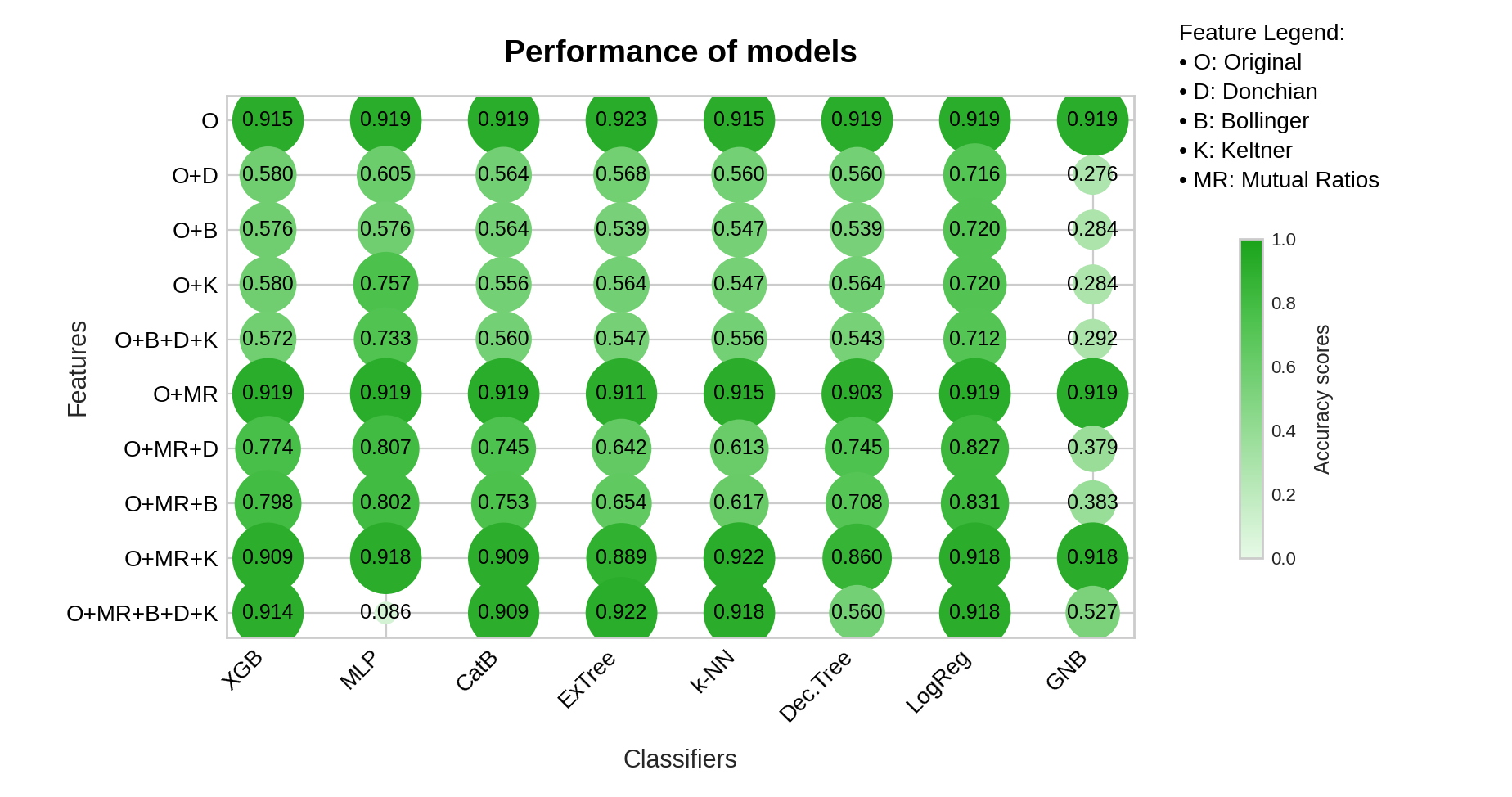}
        \caption{BSE}
    \end{subfigure}
    \caption{The plots present the accuracy scores for the \emph{open-low} across a feature-classifiers grid. Plots (a) and (b) are dedicated to S$\&$P 500 and BSE, respectively. The vertical axis shows different combinations of features, while the horizontal axis lists the classifiers. The higher the accuracy score, the better the performance. Consequently, larger circle radii and darker colors indicate better performance.}
    \label{fig:acc-low}
\end{figure*}

\begin{figure*}[h!]
    \centering
    \begin{subfigure}[b]{\textwidth}
        \includegraphics[width=0.9\textwidth]{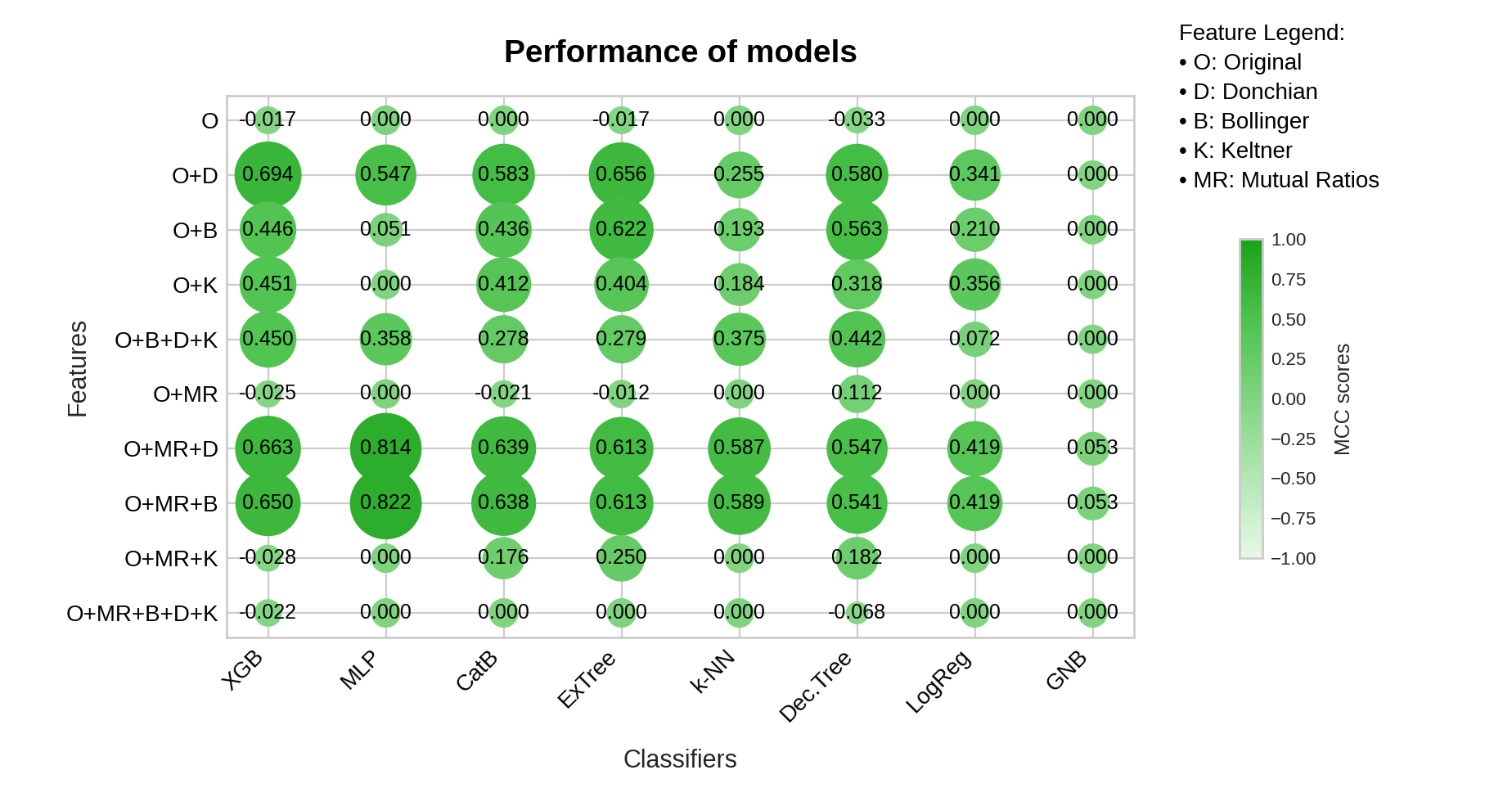}
        \caption{S$\&$P 500}
    \end{subfigure}
    \hfill
    \begin{subfigure}[b]{\textwidth}
        \includegraphics[width=0.9\textwidth]{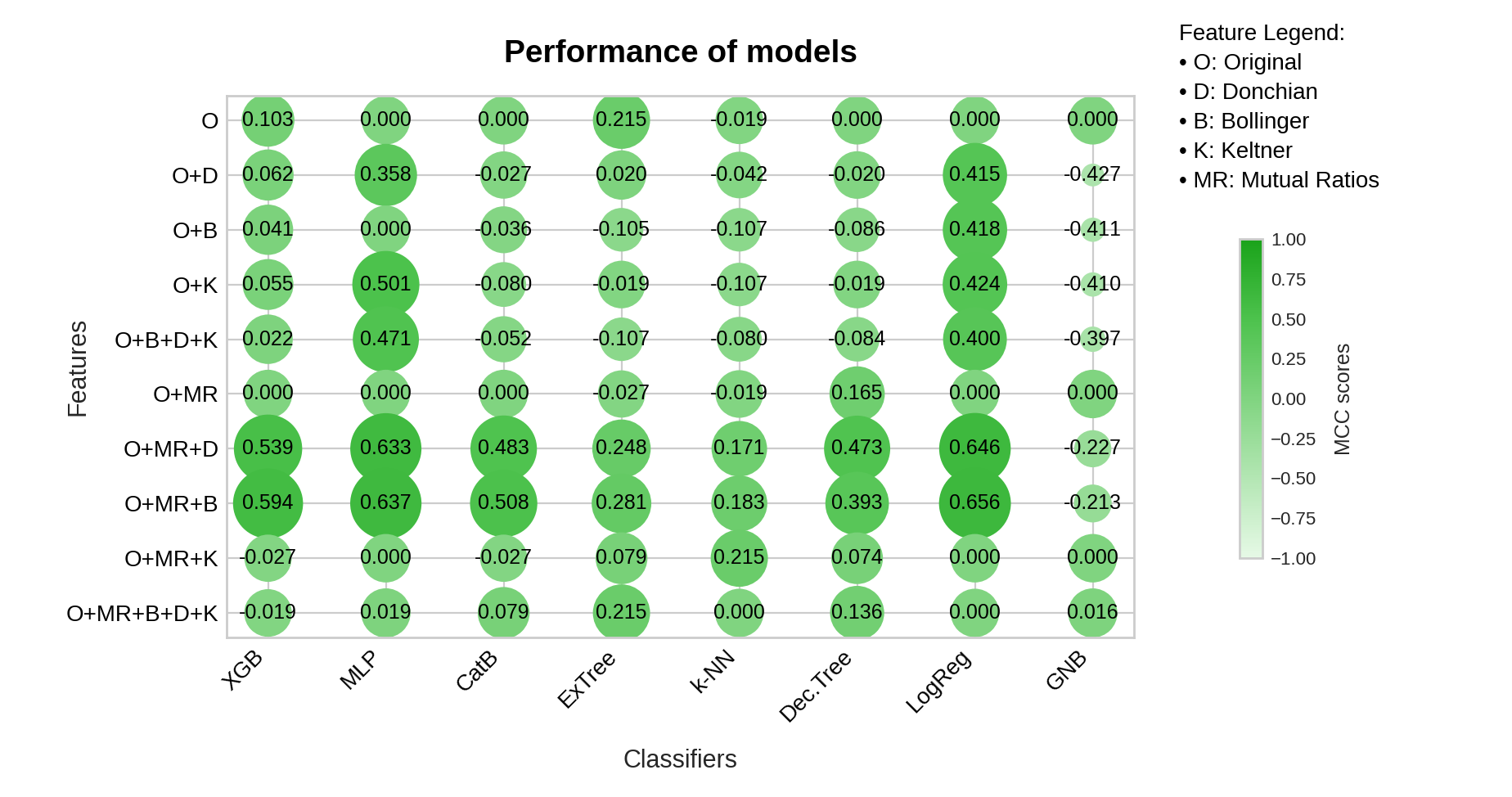}
        \caption{S$\&$P 500}
    \end{subfigure}
    \caption{The plots present the MCC scores for the \emph{open-low} across a feature-classifiers grid. Plots (a) and (b) are dedicated to S$\&$P 500 and BSE, respectively. The vertical axis shows different combinations of features, while the horizontal axis lists the classifiers. The higher the MCC score, the better the performance. Consequently, larger circle radii and darker colors indicate better performance.}
    \label{fig:mcc-low}
\end{figure*}

\item \textit{Which class of features is most effective in prediction?}

This work has explored the performance of three different genres of features and their combinations. The set of original indices constitutes the intrinsic feature set, the past volatility of the markets constitutes the historical features, and the indices' ratios of the current day (with respect to the open index) form the nowcasting features. We report results for intrinsic features and their combinations with historical and nowcasting features.

As evidenced by the figures and tabular data, the intrinsic features exhibit moderate capability to predict the rise or fall of the open index on the following day. However, the results show inconsistency in this ability across different sub-problems; the range of MCC scores, in particular, has varied substantially. The results reveal significant inconsistency in the feature’s discriminative capability across different sub-problems. The variation is particularly evident through the range of MCC scores, suggesting context-specific utility of intrinsic features. While some sub-problems exhibit strong separability (with MCC scores approaching 0.8), others show markedly weaker performance (MCC value $\tilde{0}$), indicating potential limitations in generalization capability or the influence of problem-specific factors. 

As indicated by the figures and table, the volatility components obtained from Donchian Channel, Bollinger Bands, and Keltner Channels have failed to show consistent correlation with the rise/ fall of indices. Their inclusion in the feature set has yielded contradictory results: performance improves for certain subproblems, whereas for others it is negligible or even adverse. When predicting the rise or fall of the next day's open index from the current day's high and low indices, it has shown improved performance. On the contrary, when predicting the same with respect to the current day's open index, it has shown adverse effects. Lastly, when predicting the next day's rise or fall based on the current day's close index, no measurable performance difference was observed. The cumulative results raise questions about the suitability of this class of features for predicting the rise/ fall of open index.

The results show that the addition of nowcasting features has elevated the correctness of predictions. For three of the four sub-problems, it has achieved substantial improvements; both accuracy and MCC scores have improved. The incorporation of this class of features has led to a marked improvement in the model's predictive performance. The sole exception is predicting the next day's close index, for which all features have failed to yield an admissible output. The nowcasting features have shown utility across multiple subproblems involving predicting with respect to open, high, and low indices. These findings demonstrate the ability of these features to distill meaningful economic signals for predicting the rise/ fall of open index on the subsequent trading day. 
\end{itemize}

\begin{figure*}[h!]
    \centering
    \begin{subfigure}[b]{\textwidth}
        \includegraphics[width=0.9\textwidth]{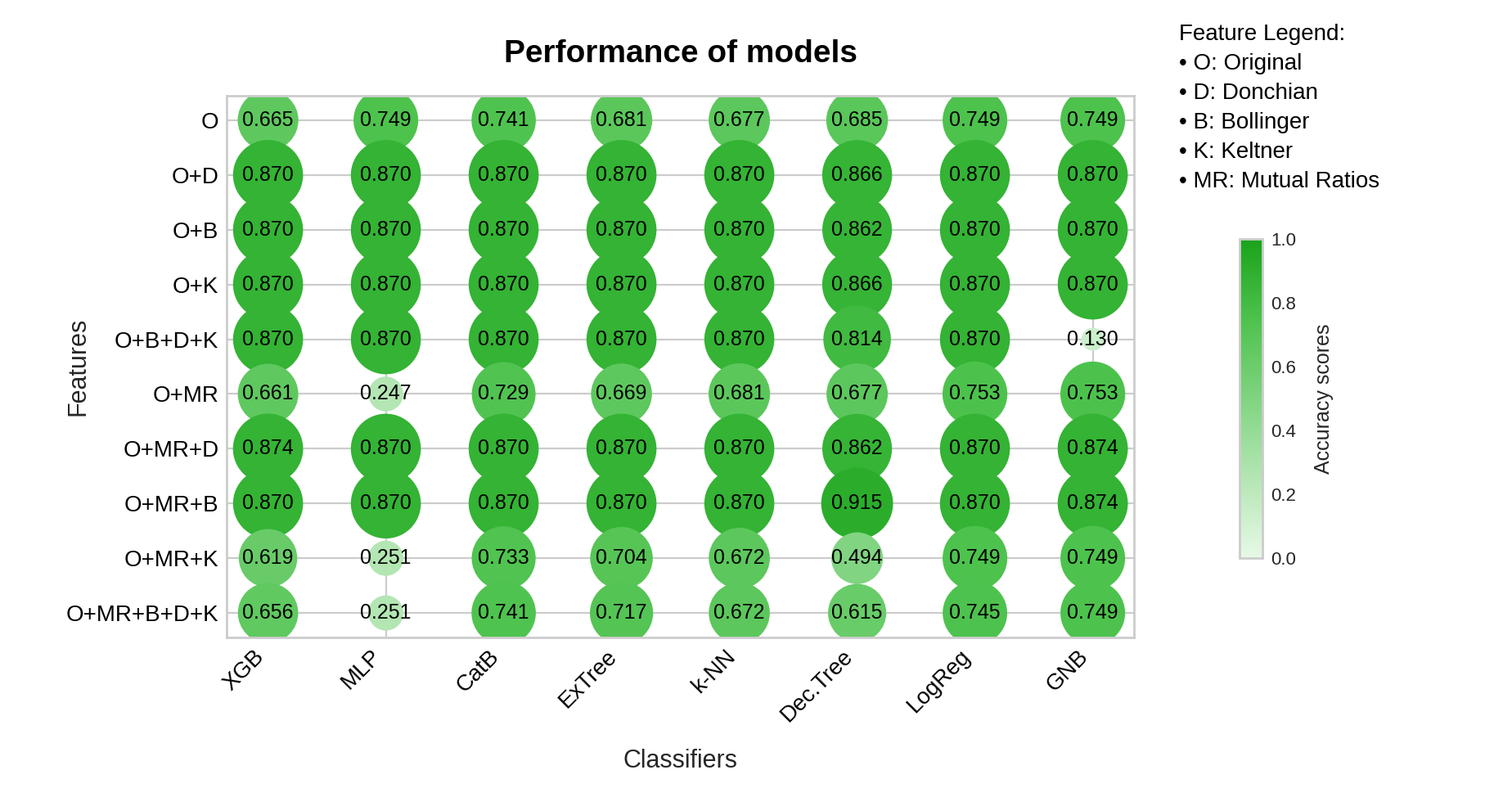}
        \caption{S$\&$P 500}
    \end{subfigure}
    \hfill
    \begin{subfigure}[b]{\textwidth}
        \includegraphics[width=0.9\textwidth]{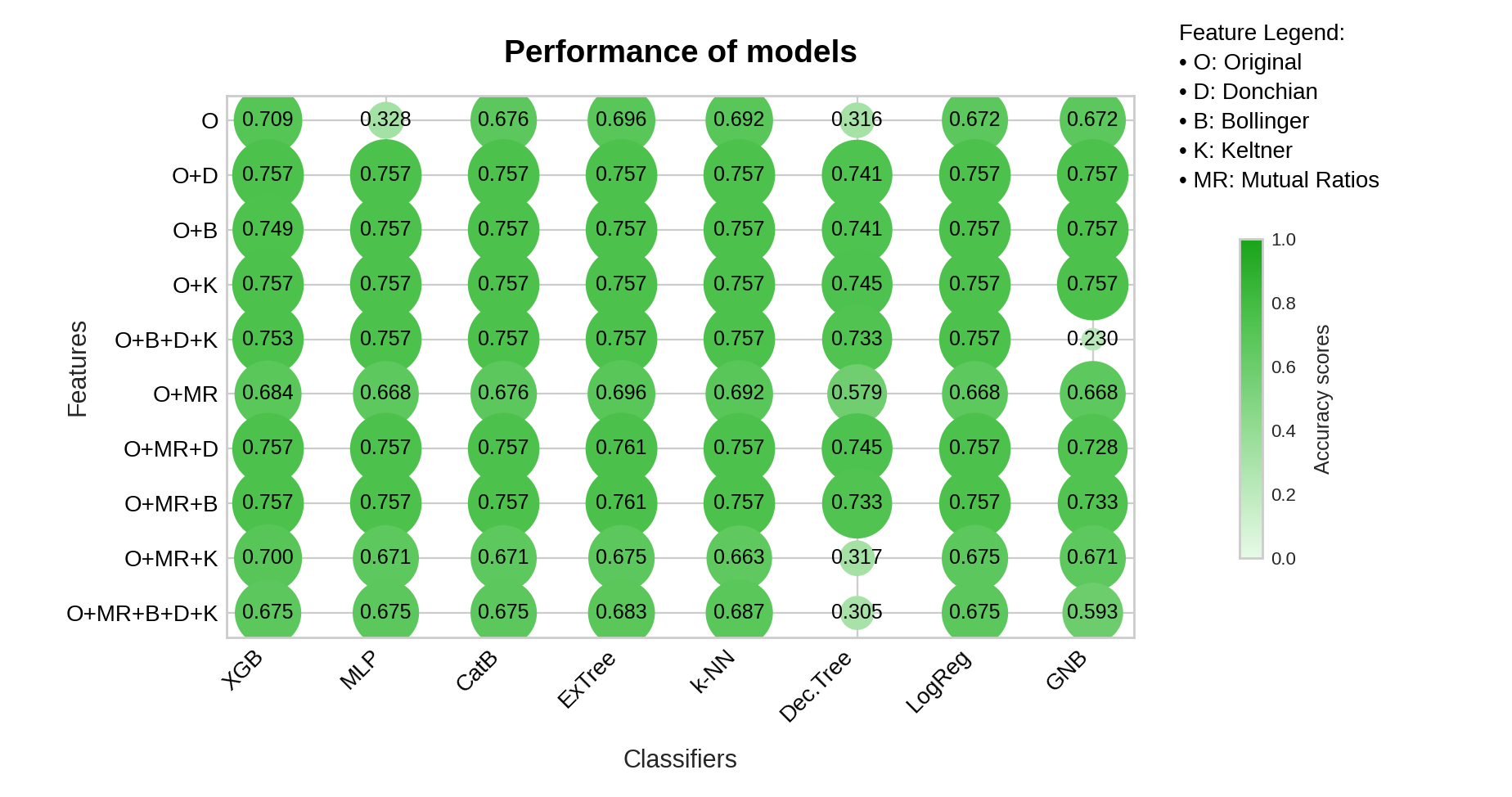}
        \caption{BSE}
    \end{subfigure}
    \caption{The plots present the accuracy scores for the \emph{open-close} across a feature-classifiers grid. Plots (a) and (b) are dedicated to S$\&$P 500 and BSE, respectively. The vertical axis shows different combinations of features, while the horizontal axis lists the classifiers. The higher the accuracy score, the better the performance. Consequently, larger circle radii and darker colors indicate better performance.}
    \label{fig:acc-close}
\end{figure*}

\begin{figure*}[h!]
    \centering
    \begin{subfigure}[b]{\textwidth}
        \includegraphics[width=0.9\textwidth]{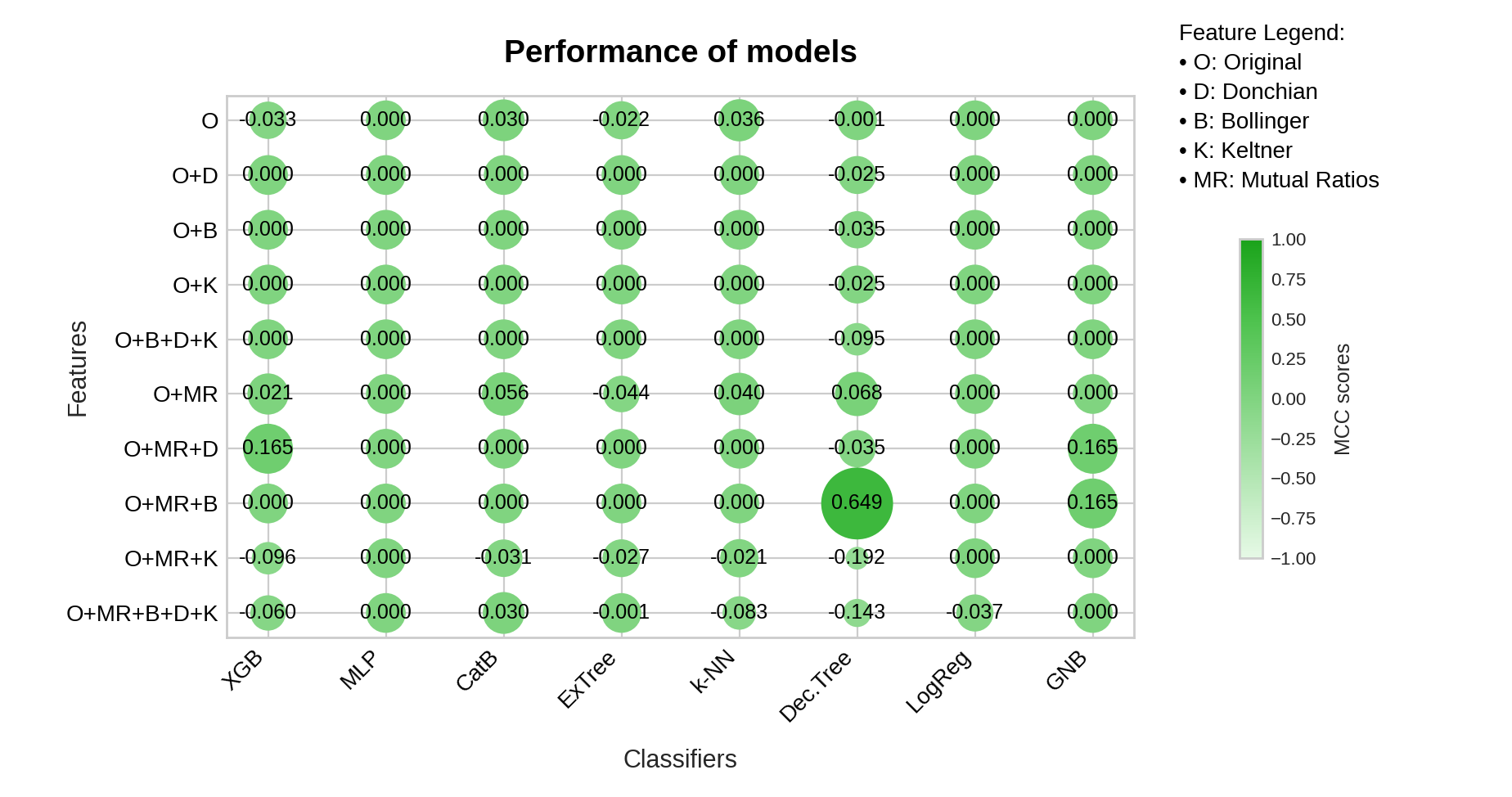}
        \caption{S$\&$P 500}
    \end{subfigure}
    \hfill
    \begin{subfigure}[b]{\textwidth}
        \includegraphics[width=0.9\textwidth]{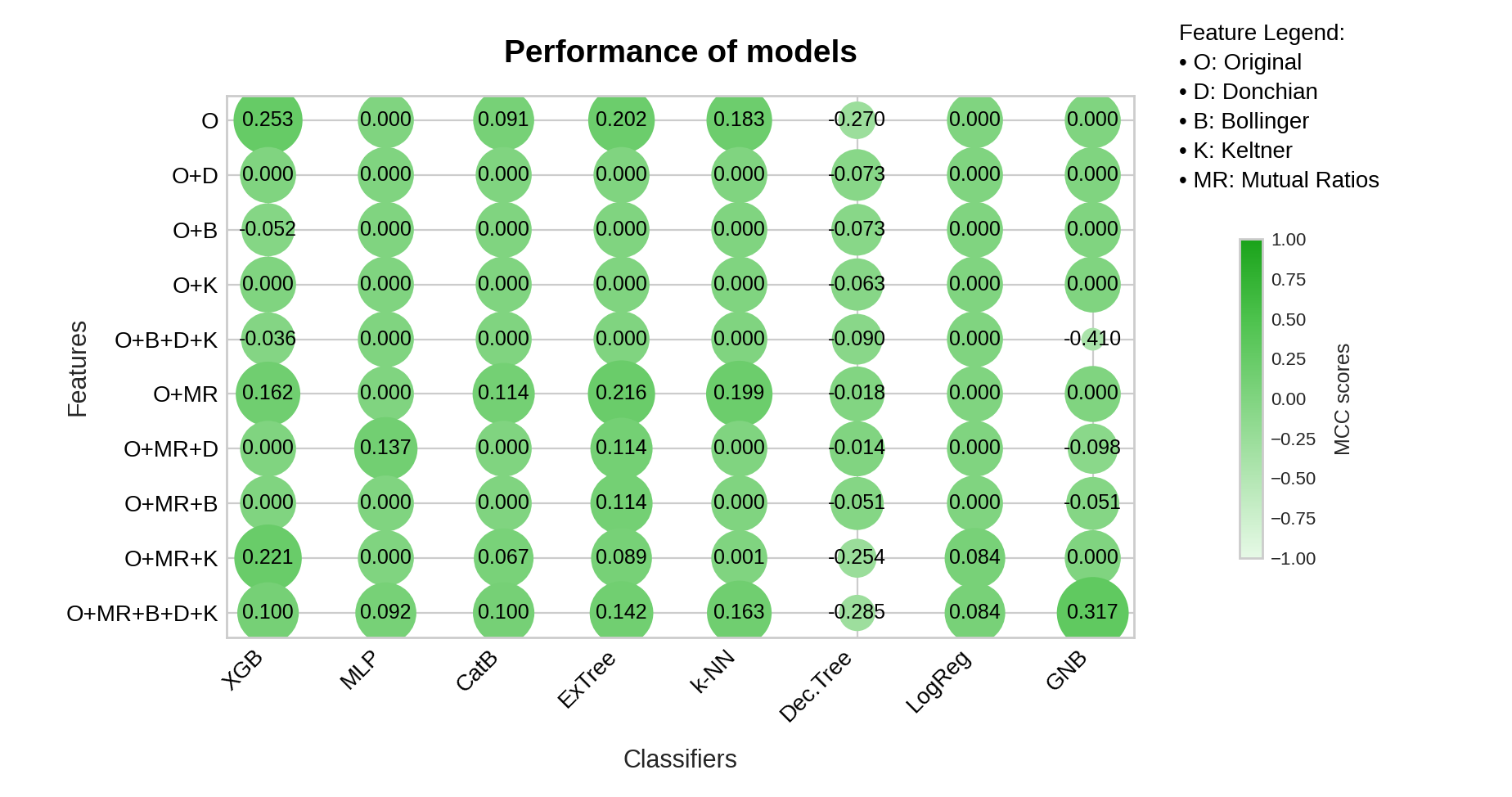}
        \caption{BSE}
    \end{subfigure}
    \caption{The plots present the MCC scores for the \emph{open-close} across a feature-classifiers grid. Plots (a) and (b) are dedicated to S$\&$P 500 and BSE, respectively. The vertical axis shows different combinations of features, while the horizontal axis lists the classifiers. The higher the MCC score, the better the performance. Consequently, larger circle radii and darker colors indicate better performance.} 
    \label{fig:mcc-close}
\end{figure*}

\begin{table*}[htbp]
\centering
\caption{Summary of reliability across different subproblems. An effective classifier is defined as achieving accuracy $\geq 0.8$ and MCC $\geq 0.65$. A checkmark ($\checkmark$) indicates that at least one such classifier exists for the subproblem; a dash (--) indicates that none met both thresholds (accuracy and MCC scores).}
\label{tab:reliability-summary}
\begin{tabular}{@{}l>{\centering\arraybackslash}p{3.8cm}*{2}{>{\centering\arraybackslash}p{1.2cm}>{\centering\arraybackslash}p{1.2cm}}@{}}
\toprule
\textbf{Task} & \textbf{Implication} & \multicolumn{4}{c@{}}{\textbf{S\&P 500} \quad \textbf{BSE}} \\
\cmidrule(lr){3-4} \cmidrule(lr){5-6}
 & & \multicolumn{2}{c}{Acc.} & \multicolumn{2}{c}{MCC} \\
\midrule
Open w.r.t. Open & Tomorrow's open $>$ today's open? & $\checkmark$ & $\checkmark$ & $\checkmark$ & $\checkmark$ \\
Open w.r.t. High & Tomorrow's open $>$ today's high? & $\checkmark$ & $\checkmark$ & $\checkmark$ & $\checkmark$ \\
Open w.r.t. Low & Tomorrow's open $>$ today's low? & $\checkmark$ & $\checkmark$ & $\checkmark$ & $\checkmark$ \\
Open w.r.t. Close & Tomorrow's open $>$ today's close? & $\checkmark$ & -- & -- & -- \\
\bottomrule
\end{tabular}
\end{table*}

\subsection{Explainability and correlation of the features with the trend}
In this subsection, we investigate i] the explainability of the features in the context of the class labels (trend prediction w.r.t. previous day's open, high, low, and close indices) and ii] the correlation of the different classes of the features with the class label. Despite using an identical feature set across tasks, their influence can vary significantly, depending on the context. To this end, we analyze feature explainability in each of the four tasks (which are described in the previous section). For each task, separate analyses are carried out for the two markets, S$\&$P 500 and BSE. 

We employ Shapley values \citep{shapley}, a game-theoretic approach, to interpret our model's predictions and understand the predictive capability of the features. This method quantifies the individual contributions of the features in a prediction task by calculating their average marginal contribution with respect to all possible combinations of features. The method's modus operandi also allows us to have a transparent and interpretable reckoning of the model's action. The magnitude of a feature's Shapley value reflects the strength of its influence on a given prediction. The overall contribution of each feature is determined by aggregating its absolute Shapley values across the entire dataset. This cumulative figure reveals which features consistently exert the strongest influence on the model's outputs, as well as their complements (weak features with low class-discriminating ability).

\begin{figure*}[h!]
    \centering
    \begin{subfigure}[b]{\textwidth}
        \includegraphics[width=0.9\textwidth]{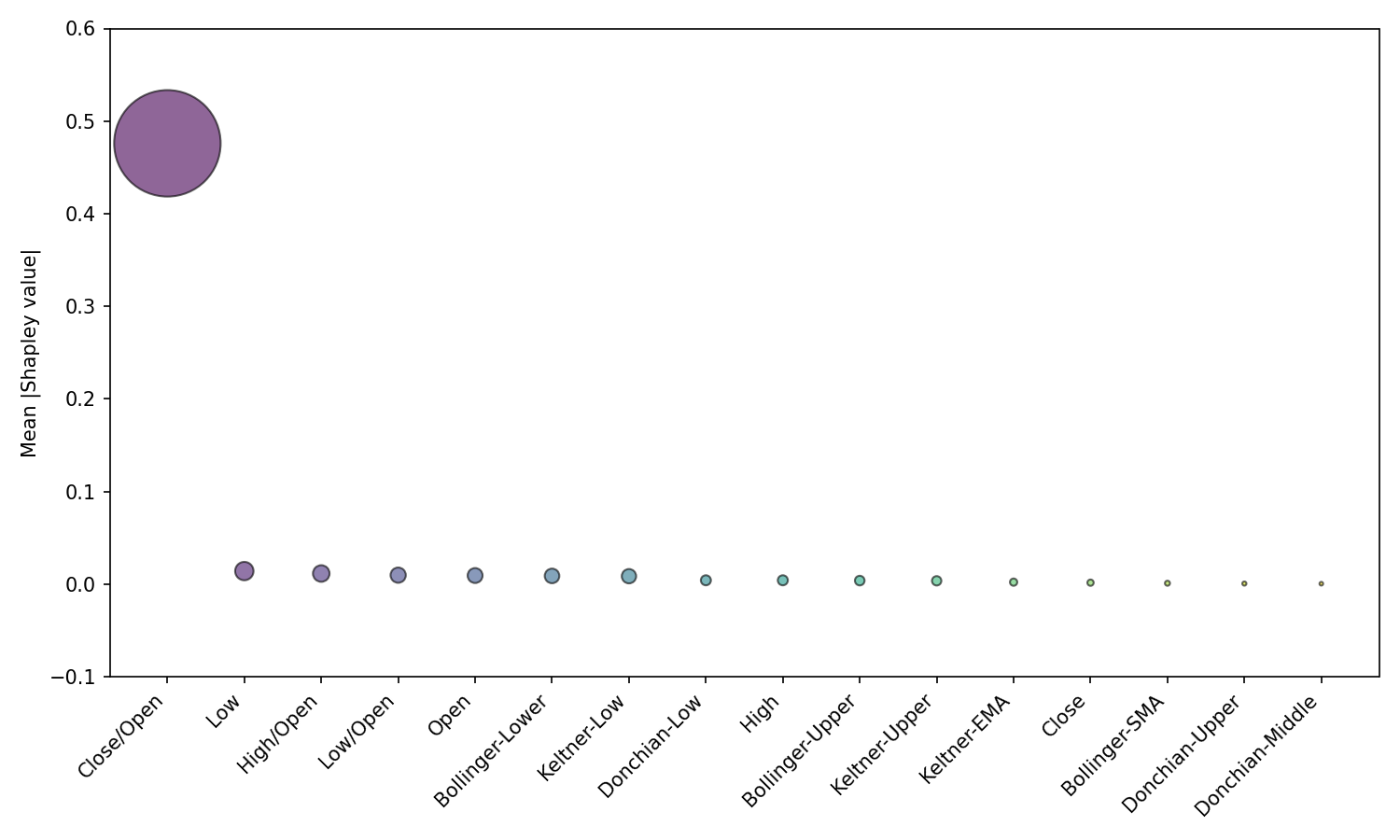}
        \caption{S$\&$P 500}
    \end{subfigure}
    \hfill
    \begin{subfigure}[b]{\textwidth}
        \includegraphics[width=0.9\textwidth]{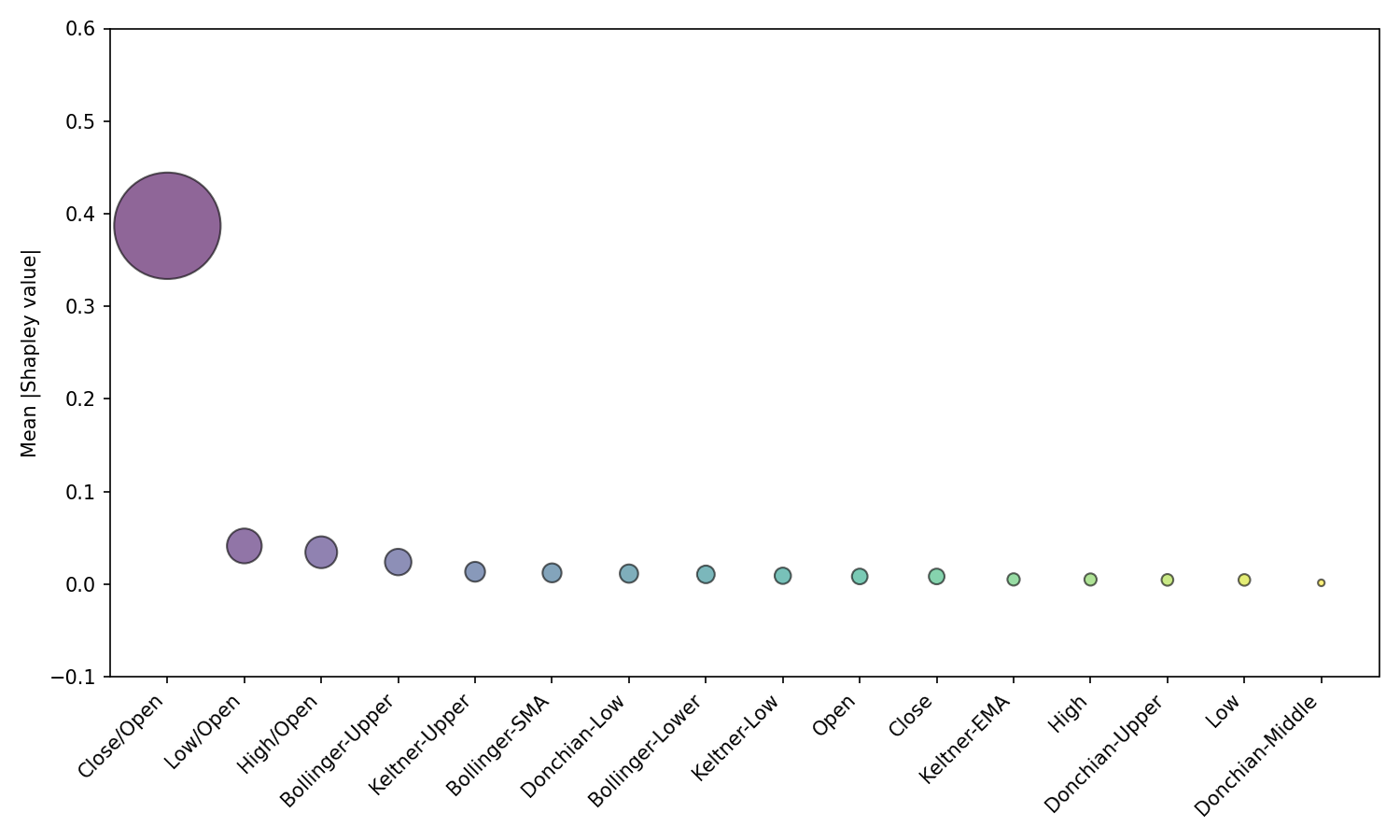}
        \caption{BSE}
    \end{subfigure}
    \caption{Shapley values obtained in the task of predicting the next day's open index's rise or fall, relative to the current day's \emph{open index}.}
    \label{fig:shap-open}
\end{figure*}

Figure \ref{fig:shap-open} shows that the ratio of the current day's close and open indices serves as the most crucial indicator in predicting the rise and fall of the open index in the upcoming trading day (with respect to the current day's open index). In both the US and the Indian markets, the remaining features exert a negligible influence on the predictions, with the effect being more pronounced in the US market. 
\begin{figure*}[h!]
    \centering
    \begin{subfigure}[b]{\textwidth}
        \includegraphics[width=0.9\textwidth]{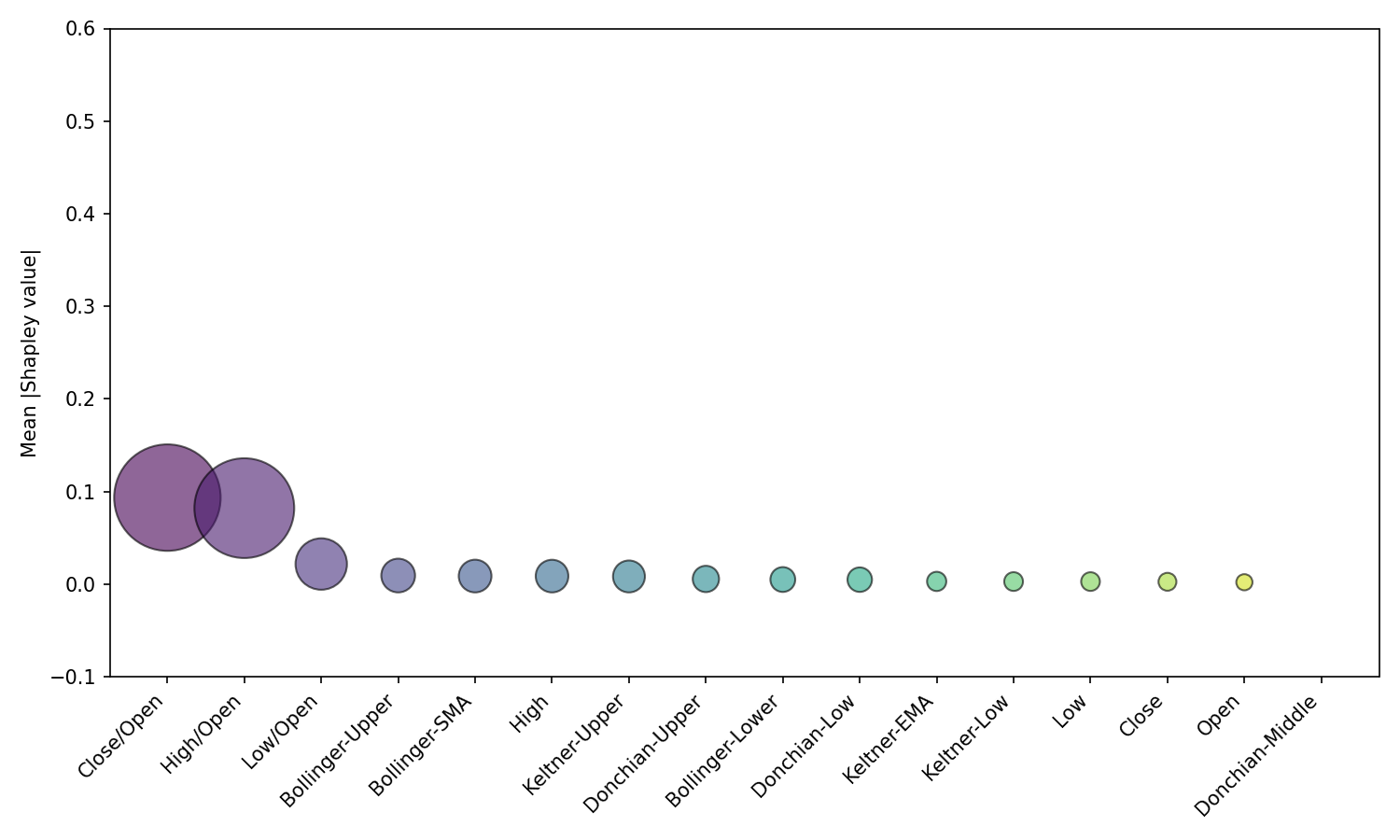}
        \caption{S$\&$P 500}
    \end{subfigure}
    \hfill
    \begin{subfigure}[b]{\textwidth}
        \includegraphics[width=0.9\textwidth]{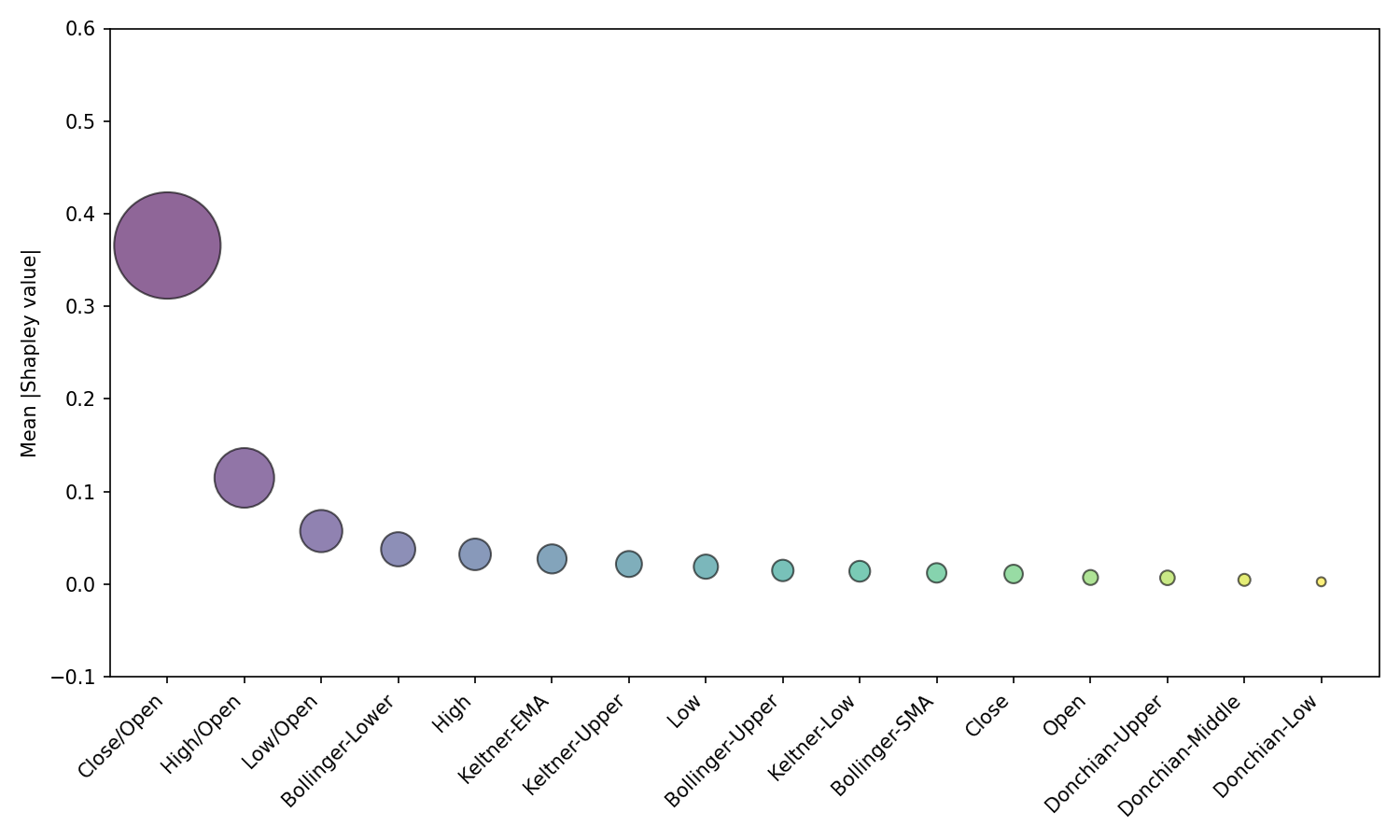}
        \caption{BSE}
    \end{subfigure}
    \caption{Shapley values obtained in the task of predicting the next day's open index's rise or fall, relative to the current day's \emph{high index}.}
    \label{fig:shap-high}
\end{figure*}
Figure \ref{fig:shap-high} shows the role of features in predicting the trend of the upcoming trading day's open index with respect to the current day's high index. While the close-to-open index ratio and the high-to-open index ratio are the most important features in both markets, their relative importance differs. In the US market, they are nearly equal. On the contrary, in the Indian market, the close-to-open index ratio is significantly more important, while the importance of the high-to-open index ratio substantially diminishes.

\begin{figure*}[h!]
    \centering
    \begin{subfigure}[b]{\textwidth}
        \includegraphics[width=0.9\textwidth]{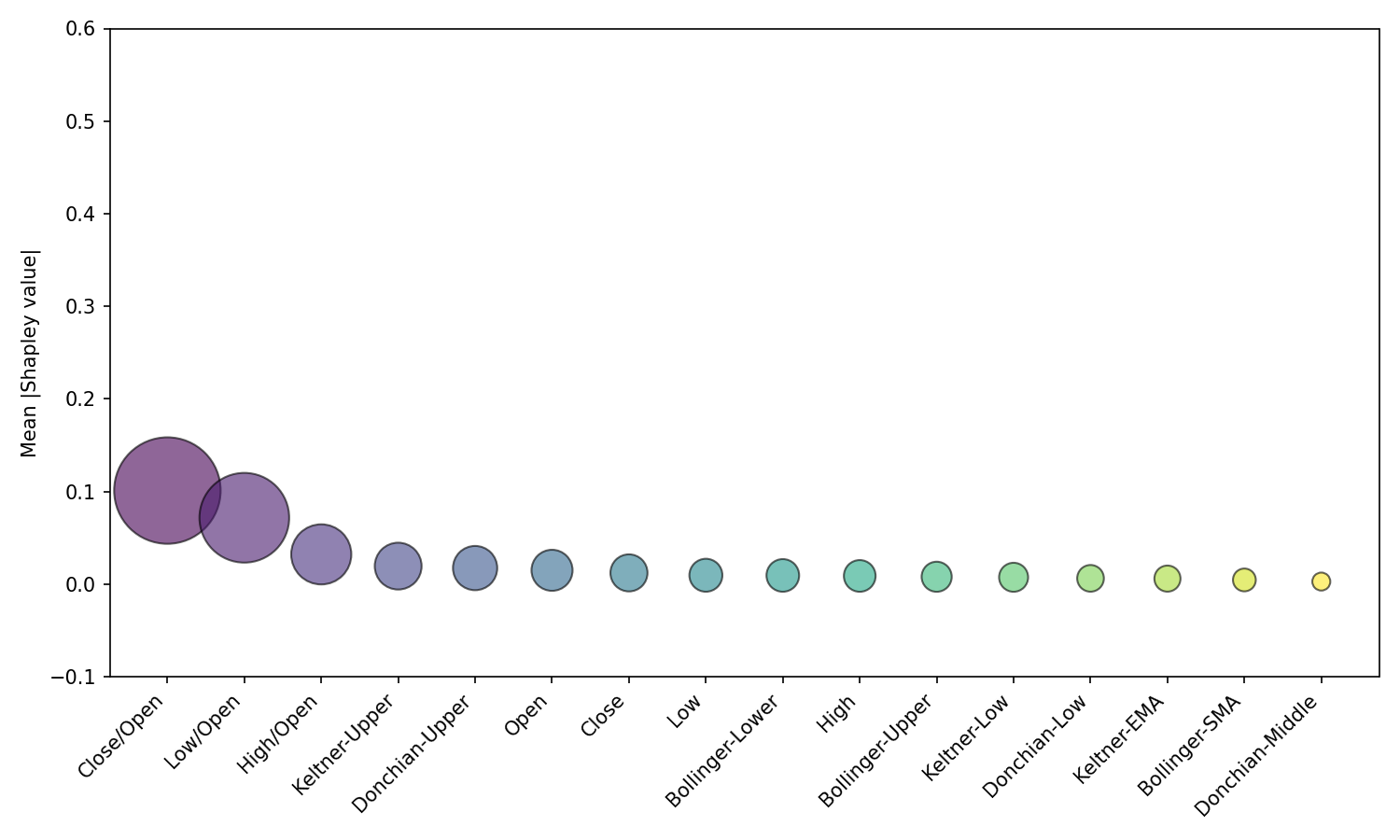}
        \caption{S$\&$P 500}
    \end{subfigure}
    \hfill
    \begin{subfigure}[b]{\textwidth}
        \includegraphics[width=0.9\textwidth]{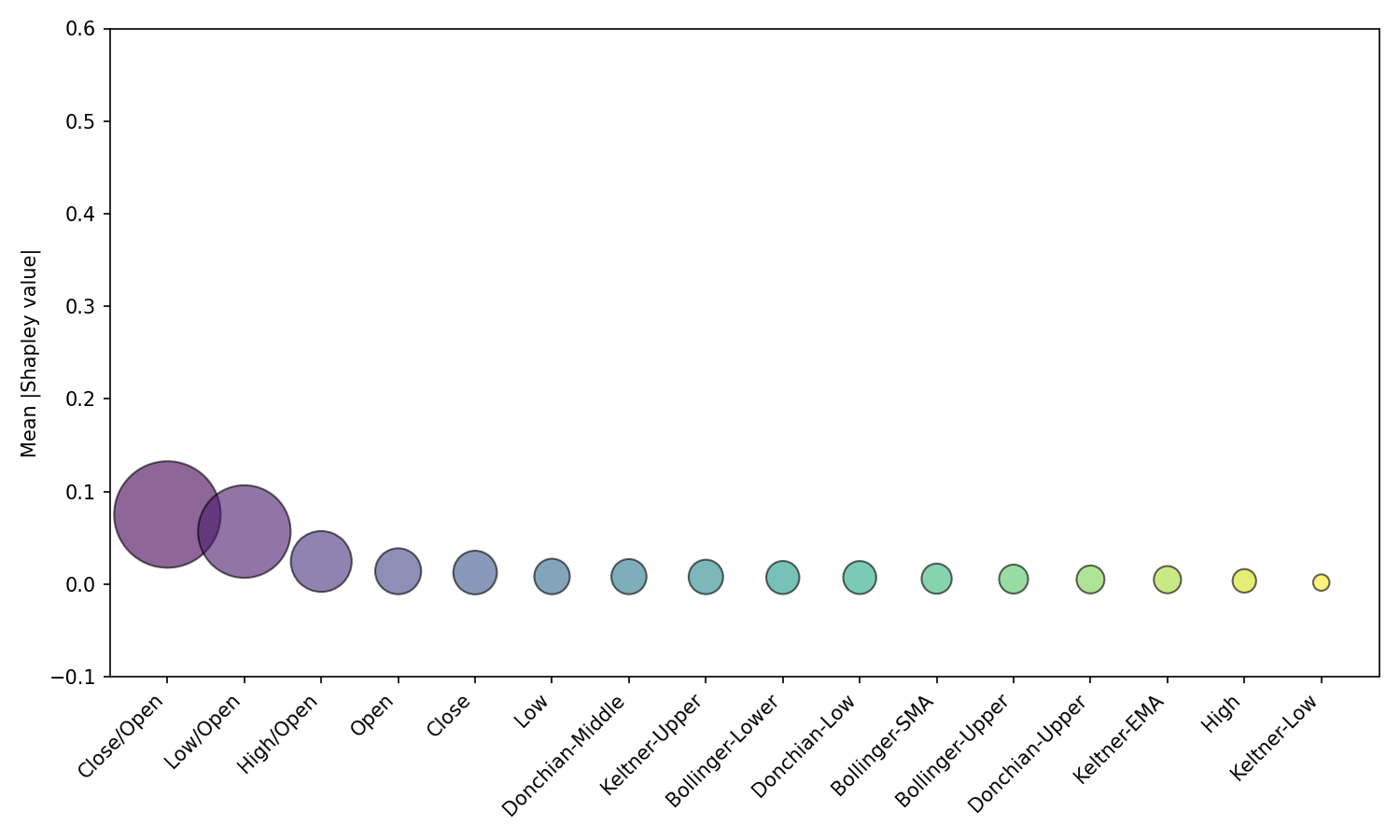}
        \caption{BSE}
    \end{subfigure}
    \caption{Shapley values obtained in the task of predicting the next day's open index's rise or fall, relative to the current day's \emph{low index}.}
    \label{fig:shap-low}
\end{figure*}
Figure \ref{fig:shap-low} shows the role of features in predicting the trend of the upcoming trading day's open index with respect to the current day's low index. For this task, the US market and the Indian market have shown nearly identical trends for the most important features. In either case, the close-to-open index ratio and the low-to-open index ratio have been identified as the most important features, with the close-to-open ratio taking a slight precedence.
\begin{figure*}[h!]
    \centering
    \begin{subfigure}[b]{\textwidth}
        \includegraphics[width=0.9\textwidth]{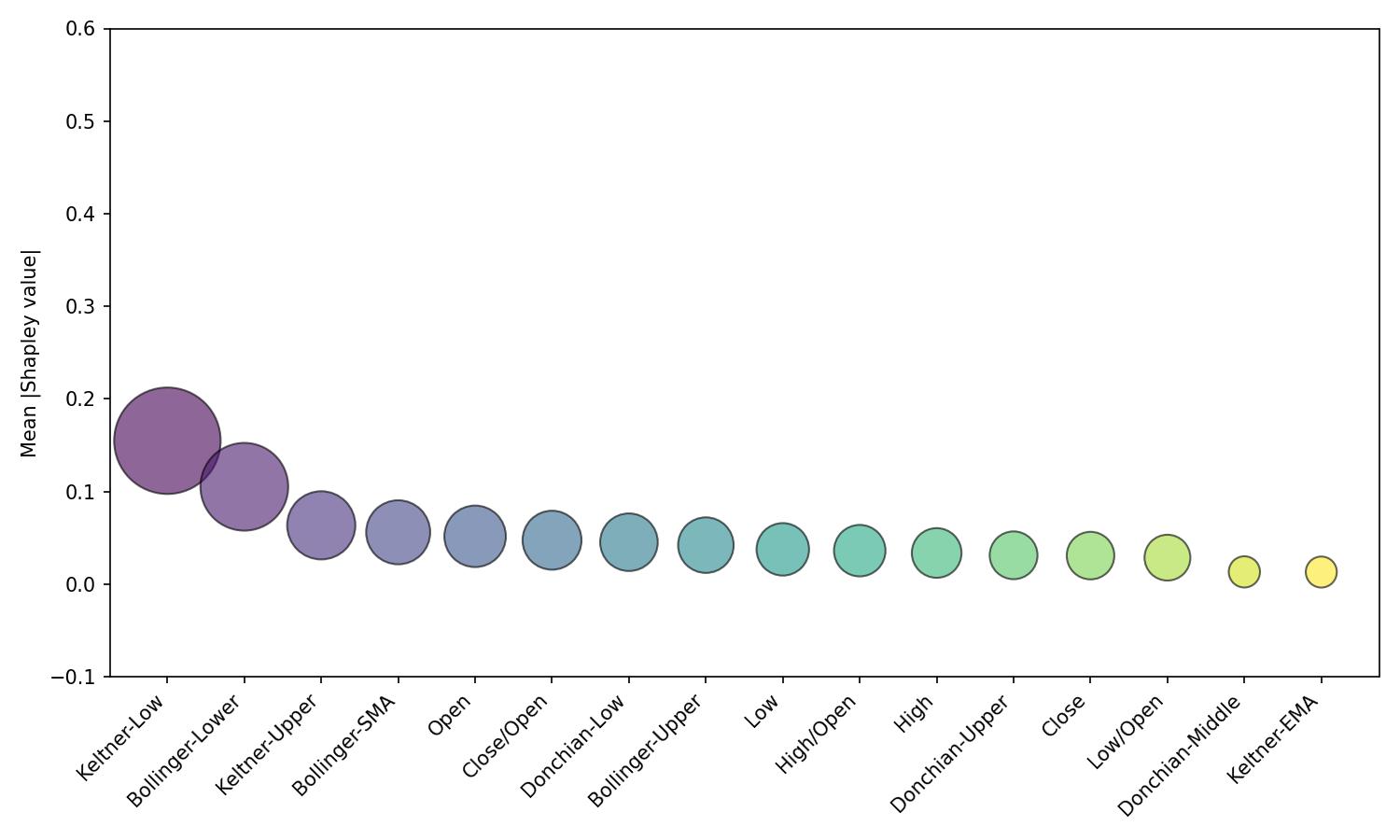}
        \caption{S$\&$P 500}
    \end{subfigure}
    \hfill
    \begin{subfigure}[b]{\textwidth}
        \includegraphics[width=0.9\textwidth]{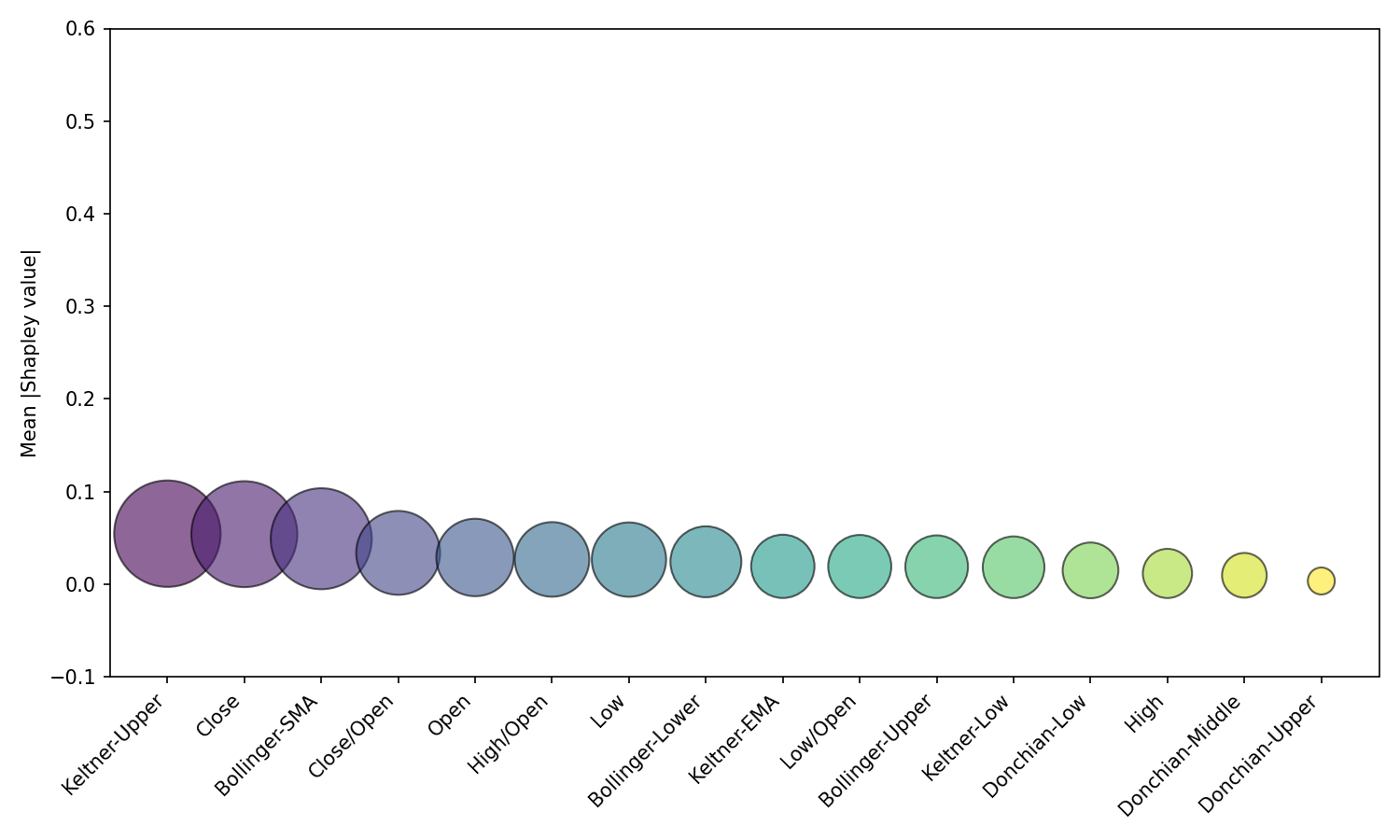}
        \caption{BSE}
    \end{subfigure}
    \caption{Shapley values obtained in the task of predicting the next day's open index's rise or fall, relative to the current day's \emph{close index}.}
    \label{fig:shap-close}
\end{figure*}
Figures \ref{fig:acc-close} and \ref{fig:mcc-close} show that the proposed modus operandi has shown sub-optimal efficacy in predicting the trend of the upcoming trading day's open index with respect to the current day's close index. The result is in contradiction with the performance achieved in the other three sub-problems, for which we achieved an admissible performance.
This performance discrepancy is reflected and in congruence in the feature importance analysis shown in \ref{fig:shap-close}. For the US Market, the lower band of the Keltner Channel emerges as the most important feature, followed by the lower band of the Bollinger Bands. However, the overall importance values are low—the highest is only 0.15, and the narrow range across features indicates a lack of clear discriminatory power. The scenario is more diffuse in the Indian market. The upper band of the Keltner channel, the closing index of the current day, and the moving average of Bollinger Bands emerge as the top contributors, but the importance values are exceptionally low (max < 0.1) and exhibit a near-random distribution. The consistently low feature importance scores across both markets indicate that the available features lack the strong correlations with the class label necessary for effective prediction of the rise-or-fall of the upcoming trading day's open index with respect to the current day's close index.
\section{Discussion}
This research challenges the apparent randomness of the healthcare sector's index and seeks to uncover the underlying factors that contribute to its rise/ fall. The fundamental features that are associated with the index's directional movement remain poorly understood \citep{unpredictable1,unpredictable2}, presenting a significant knowledge gap that this study aims to address. We consider a diversified class of factors to identify those that strongly correlate with the directional movement at market opening and can serve as effective features for modeling the problem. The three types considered are: i] raw index values, ii] historical movement, and iii] nowcasting factors. The results of classifier modeling and feature importance assessment have shown that historical price movement has negligible power in accurately predicting the rise or fall of the opening value of the healthcare index. On the contrary, nowcasting factors or the index's movement in the prior day (of prediction) show a stronger association with the rise or fall of the index, thereby rendering its correct prediction. The original index values have demonstrated a certain degree of predictive power, yielding acceptable accuracy and MCC scores on several occasions. Experimental results also show that the combination of these two classes of data (nowcasting factors and raw index values) can serve as an effective feature for predicting the indices' movement.

To ensure the viability and robustness of our exploration, we have conducted the study using data from two different economies: the US and India. We have obtained mostly similar outcomes in the two scenarios; sectoral homogeneity is a contributing factor in this regard. The predictive performance for the US market is slightly better than that of the Indian market, likely due to the greater stability and higher trading volume of the US market \footnote{\url{https://www.zyen.com/publications/public-reports/the-global-financial-centres-index-37/}} \footnote{\url{https://www.imf.org/en/Publications/WEO/weo-database/2024/April/groups-and-aggregates}}. This stability strengthens the association between features and market movements, increasing the veracity of the predictions. This study is valuable not only to researchers but also to stakeholders across various domains, including finance and healthcare.

\subsection{Implications in the healthcare and financial sectors}
The growth of the healthcare index is closely tied to breakthroughs and developments in the sector: an innovation raises the index, while a setback lowers it. Conversely, a rise in the index can boost fund flows for innovation and development in the sector, while a fall in the index often takes a toll on investor confidence, thereby reducing fund flows. Together, through mutual feedback, they establish a bidirectional causality. In this context, prior knowledge of the rise or fall of the sectoral index can help investors obtain timely financial clues to allocate their funds. The credible clues can build confidence in investors' perceptions of the sector's stability and growth. Being able to predict the market can enhance one's understanding, leading them to make informed decisions about their fund allocations and thereby increasing the flow of funds. Our model's features are derived from publicly available healthcare index data, making them equally accessible to all. This transparency helps counteract biases that can stem from exclusive access to information (by influential entities). Transparency can prevent insider trading, which is known to have a disastrous effect on finances through capital misallocation and erosion of investor confidence \citep{informed-trading1, insider-trading2}. In the healthcare sector, insider trading can have disastrous effects, like retarded milestones in the health sector by affecting innovations like new drug trials, the launch of new drugs, and vaccines.

\subsection{Limitations and directions of future research}
Despite its contributions, this study has limitations that suggest directions for future research. Firstly, the results of predicting the upcoming trading day's rise-or-fall of the open index with respect to the current day's close index are sub-optimal. Two primary factors can explain this outcome: either the chosen features lack association with the rise-or-fall of the upcoming trading day's open index with respect to the current day's close index, or the rise-or-fall of the upcoming trading day's open index with respect to the current day's closing index itself exhibits a random nature. 

The second limitation arises when forecasts serve as a feature in themselves. Before deploying our framework for rise-or-fall prediction, we must consider the following: the act of forecasting can actively shape the future it attempts to foresee. This is commonly referred to as the observer effect, where predictions given by a trusted model become new input to the system. For example, a predicted economic downturn can encourage policy changes that prevent it, or a forecast of high demand can trigger panic buying that creates it. This renders a non-static model validity, which is contingent on how its predictions are integrated into the decision-making ecosystem. As our features and model are crafted from publicly available data, we must be cautious about how its predictions can influence the actual market, particularly if it is adopted on a large scale.

\section{Conclusion}

This research challenges the prevailing perception that short-term movements in healthcare sector indices are mostly random and difficult to predict. Through a systematic investigation of multiple genres of features — raw index values, technical tool-based historical price movement, and nowcasting factors — we demonstrate that an admissible and economically meaningful signal is camouflaged in the apparent noise in daily healthcare index fluctuations. We frame the problem as a supervised classification task involving a one-step-ahead rolling window. 

Our empirical findings consistently reveal that nowcasting features, constructed from mutual ratios of open–high–low–close indices in the current trading day, are the dominant contributors to predictive performance at the upcoming trading day. In contrast, historical price movement exhibits negligible predictive association with the directional movement of the opening index, while raw index values offer moderate but inconsistent predictive utility. Importantly, integrating nowcasting features with raw price information yields the most reliable performance, indicating that near-real-time market signals complemented by baseline price level fluctuation provide the most effective representation of short-term healthcare index behavior. Shapley-based feature explainability analysis also reveals the same. The robustness of these findings is validated across two structurally distinct healthcare markets—the United States and India. The slightly stronger predictive performance observed in the U.S. market can plausibly be attributed to greater market depth, liquidity, and stability, which reinforce the association between observed features and subsequent price movements.

Beyond methodological contributions, this research has important implications for both healthcare ecosystems. Short-term predictability of healthcare indices can support more efficient capital allocation, stabilize funding flows for medical research and development, and reduce information asymmetry among market participants of different classes. Our methodology relies exclusively on publicly available data; the proposed framework promotes transparency and equitable access to market intelligence, thereby mitigating the risks of insider trading, which can undermine innovation and question trust in healthcare financing systems.

This work is not devoid of limitations and ethical considerations. One predictive task — the subproblem related to open and close indices — remains underserved. The cause can be intrinsic randomness in the scenario or missing explanatory factors for the subproblem. Additionally, forecasts can influence market behavior. In future research, we will examine the feedback effects of rendering predictions and the systematic implications of deploying predictive models at scale. We will address these interdisciplinary challenges by integrating behavioral, policy, and structural factors into future modeling.

\bibliographystyle{elsarticle-num-names}
\bibliography{cas-refs}

\appendix
\section{Volatility Computation}
\label{sec:vol}  
Volatility gives a normalized measure of index fluctuations over a period. The initial step involves the estimation of the directional and magnitude changes at Day $t$, denoted by $r(t)$, by taking the logarithm of the consecutive indices $P(t)$ and $P(t-1)$. The mean change is computed through the averaging of $r(t)$ over the days in question, followed by computing the variance, which measures the degree of dispersion around this mean. Volatility is obtained by taking the square root of the variance. For computing period-specific volatility (essential for comparing assets across different time frames), daily volatility is weighted by the square root of the number of trading days. 
\subsection{1. Logarithmic Returns}
Given a index series \( P_t \), the logarithmic return at time \( t \) is:
\begin{equation}
r(t) = \ln\left(\frac{P(t)}{P({t-1})}\right)
\end{equation}

\subsection{2. Mean Return}
The average return over \( N \) periods:
\begin{equation}
\bar{r} = \frac{1}{N} \sum_{t=1}^{N} r(t)
\end{equation}

\subsection{3. Variance of Returns}
The sample variance (unbiased estimator):
\begin{equation}
\sigma^2 = \frac{1}{N-1} \sum_{t=1}^{N} (r(t) - \bar{r})^2
\end{equation}

\subsection{4. Volatility (Standard Deviation)}
Daily volatility is the square root of variance:
\begin{equation}
\text{Volatility}_{\text{daily}} = \sqrt{\sigma^2}
\end{equation}

\subsection{5. Periodized Volatility}
Scaled by the square root of time (assuming 252 trading days/year):
\begin{equation}
\text{Volatility}_{\text{periodized}} = \text{Volatility}_{\text{daily}} \times \sqrt{\text{No. of days}}
\end{equation}

\section{Accuracy}
\label{sec:acc}
For assessing a classification task, performance is quantified using counts derived from a confusion matrix. The Accuracy (\(acc\)) metric represents the overall proportion of correct predictions and is formally defined as:

\[
acc = \frac{TP + TN}{TP + TN + FP + FN}
\]

where:
\(TP\) = True Positives,
\(TN\) = True Negatives,
\(FP\) = False Positives,
\(FN\) = False Negatives.

\section{Matthews Correlation Coefficient (MCC) scores}
\label{sec:mcc}
It is a robust metric for evaluating binary classification models, effective across varying distributions of classes. The MCC score is high only when a model produces efficient performance for all possible classes.

MCC is calculated as:

\[
\text{MCC} = \frac{TP \times TN - FP \times FN}{\sqrt{(TP + FP)(TP + FN)(TN + FP)(TN + FN)}}
\]

For a binary classification problem, an accuracy of $0.5$ is the standard expectation from a random classifier. In this research, we deal with a binary classification task. Accordingly, we set the threshold at $0.8$ to consider the model to be an effective one.

\subsection{Interpretation of MCC score Ranges}
MCC scores can range from (\(-1 \leq \text{MCC} \leq +1\)). \(\text{MCC} = +1\) signifies perfect prediction (all correct, no errors). \(\text{MCC} \geq 0.7\) indicates a very strong correlation between ground truth and predictions (excellent model).
\(0.5 \leq \text{MCC} < 0.65\): shows strong correlation between actual labels and predictions (good model).
\(0.3 \leq \text{MCC} < 0.5\): indicates a moderate performance.
\(0.1 \leq \text{MCC} < 0.3\): signifies a significantly weak performance with low correlation between ground truth and predictions (poor model).
\(\text{MCC} < 0.1\): indicates very feeble correlation (unreliable).
\(\text{MCC} = 0\): signifies no correlation and random prediction. This occurs when the model's correct predictions (\textit{TP, TN}) are statistically indistinguishable from its errors (\textit{FP, FN}), equivalent to random guessing.
\(\text{MCC} = -1\): signifies a flipping of predictions (always wrong). Consequently, we have chosen $\text{MCC}=0.65$ as a threshold to identify efficient models.

\end{document}